\documentclass[sn-mathphys-num]{sn-jnl}

\usepackage{xfrac}
\usepackage{xurl}
\usepackage{hyperref}
\usepackage{booktabs}

\usepackage{tikz}
\usepackage{tabularx}
\usepackage{listings}
\usepackage{nicefrac}
\usepackage{multirow}
\usepackage{hhline}
\usepackage{graphicx}
\usepackage{calc}
\usepackage{array}
\usepackage{soul}
\usepackage{proof}
\usepackage[export]{adjustbox}
\usepackage{verbatim}
\usepackage{stmaryrd}
\usepackage{amsfonts}
\usepackage{mathtools}
\usepackage{caption}
\usepackage{subcaption}
\usepackage[framemethod=TikZ]{mdframed}
\usepackage[inline,shortlabels]{enumitem}
\setlist[enumerate,1]{label={(\arabic*)}}
\usepackage[normalem]{ulem}
\usepackage{transparent}
\usepackage{arydshln}
\usepackage{makecell}
\usepackage{amsmath}
\usepackage{amsthm}
\usepackage{amssymb}
\usepackage{textcomp}
\usepackage{verbatimbox}
\usepackage{wasysym}

\usepackage{fancyvrb}
\usepackage{comment}

\usepackage{algorithm} 
\usepackage{algpseudocode} 

\definecolor{keywordcolor}{rgb}{0.5,0,0.1}
\definecolor{stringcolor}{rgb}{0,0,1}
\definecolor{emphColor}{rgb}{0,0,0}
\definecolor{codecommentcolor}{rgb}{0.03,0.6,0}
\definecolor{codegreen}{rgb}{0,0.6,0}
\definecolor{codegray}{rgb}{0.5,0.5,0.5}
\definecolor{codepurple}{rgb}{0.58,0,0.82}
\definecolor{mediumgray}{rgb}{0.80, 0.80, 0.80}
\definecolor{listinggray}{rgb}{0.9,0.9,0.9}

\lstdefinelanguage{vql}{
    morekeywords={@QueryBasedFeature,@Constraint,count,pattern,private,neg,find,import,true,false,or,check,job,action,state,severity,
    message,oclIsKindOf,self,exists,includes,invariant,class},
    sensitive=true, 
    morecomment=[l]{//},
    morecomment=[s]{/*}{*/},
	  morestring=[b]{"},
}

\lstdefinestyle{mystyle}{
  breaklines=true,
  showstringspaces=false,
  basicstyle=\ttfamily\scriptsize,
  identifierstyle=\color{black},
  stringstyle=\color{orange},
  columns=flexible,
  keywordstyle=[0]*\bfseries\color{blue!70!black},
  keywordstyle=[1]\bfseries\color{blue!70!black}, 
  commentstyle=\color{codecommentcolor},
  backgroundcolor=\color{black!2},
  frame=single, 
  numbers=left,
  numbersep=5pt,
  numberstyle=\ssmall\color{gray},
  captionpos=b,
  xrightmargin=3pt,
  xleftmargin=3pt,
  keepspaces=true,
  showspaces=false,
  breakatwhitespace=false,
  tabsize=2,
  showtabs=false,
}
\surroundwithmdframed[
  topline=false,
  rightline=false,
  bottomline=false,
  leftmargin=\parindent,
  skipabove=\medskipamount,
  skipbelow=\medskipamount,
  linewidth=4pt,
  linecolor={black},
  backgroundcolor={black!5},
  leftmargin=0cm
]{proposition}

\surroundwithmdframed[
  topline=false,
  rightline=false,
  bottomline=false,
  leftmargin=\parindent,
  skipabove=\medskipamount,
  skipbelow=\medskipamount,
  linewidth=4pt,
  linecolor={black!50},
  backgroundcolor={black!5},
  leftmargin=0cm
]{definition}

\surroundwithmdframed[
  topline=false,
  rightline=false,
  bottomline=false,
  leftmargin=\parindent,
  skipabove=\medskipamount,
  skipbelow=\medskipamount,
  linewidth=4pt,
  linecolor={gray!50},
  leftmargin=0cm
]{example}

\newmdenv[
  linewidth=1pt,
  roundcorner=5pt 
]{findingRuled}

\newmdenv[
  topline=false,
  bottomline=false,
  rightline=false,
  skipabove=\topsep,
  skipbelow=\topsep,
  leftmargin=-10pt,
  rightmargin=-10pt,
  innertopmargin=6pt,
  innerbottommargin=6pt,
  linewidth=4pt,
  linecolor={blue!50},
  backgroundcolor={blue!5},
]{thesisQuestionRuled}

\newmdenv[
  topline=false,
  bottomline=false,
  rightline=false,
  skipabove=\topsep,
  skipbelow=\topsep,
  leftmargin=-10pt,
  rightmargin=-10pt,
  innertopmargin=6pt,
  innerbottommargin=6pt,
  linewidth=4pt,
  linecolor={black},
  backgroundcolor={black!5},
]{thesisResultRuled}

\newmdenv[
  topline=false,
  bottomline=false,
  rightline=false,
  skipabove=\topsep,
  skipbelow=\topsep,
  leftmargin=-10pt,
  rightmargin=-10pt,
  innertopmargin=6pt,
  innerbottommargin=6pt,
  linewidth=4pt,
  linecolor={orange!80!red},
  backgroundcolor={orange!5},
]{hypothesisRuled}

\definecolor{viatraEmphColor}{RGB}{0,80,125}
\definecolor{keywordcolor}{rgb}{0.5,0,0.1}
\definecolor{commentcolor}{rgb}{0,0.3,0.1}
\definecolor{stringcolor}{rgb}{0,0,1}

\lstdefinelanguage{viatra}
{
morekeywords={@QueryBasedFeature,@Constraint,count,pattern,package,neg,find,import,true,false,or,check,job,action,state,severity,location,message,oclIsKindOf,self,exists,includes,invariant,class,private,epackage,java},
emph={Pseudostate,Vertex,Region,Transition,Entry,Synchronization,State,RegularState,CompositeElement,Trigger,Guard,Action,Statechart,vertices,regions,source,target,incomingTransitions,trigger,guard,action,elements,newElements,PartialModel,open,must,may,var,element,transitions,FamilyTree,members,Member,parents,age,VisionBlocked,blockedBy,Actor,xPos,yPos,ySpeed,placedOn,Lane_Horizontal,length,width},
emphstyle={\color{viatraEmphColor}},
sensitive=true, morecomment=[l]{//}, morecomment=[s]{/*}{*/},
morestring=[b]{"}
}
\lstdefinestyle{viatrasmall}{
	basicstyle=\scriptsize\ttfamily,
	commentstyle=\color{commentcolor}\ttfamily,
	stringstyle=\color{stringcolor}\ttfamily,
	captionpos=b,
	keywordstyle=\color{keywordcolor}\bfseries\ttfamily,
	showstringspaces=false,
	tabsize=2,
	language=viatra,
	escapeinside={(*@}{@*)}
}
\lstdefinestyle{viatrabig}{
	basicstyle=\ttfamily,
	commentstyle=\color{commentcolor}\ttfamily,
	stringstyle=\color{stringcolor}\ttfamily,
	captionpos=b,
	keywordstyle=\color{keywordcolor}\bfseries\ttfamily,
	showstringspaces=false,
	tabsize=2,
	language=viatra,
	escapeinside={(*@}{@*)}
}

\definecolor{emphColor}{rgb}{0.1,0.1,0.1}  

\definecolor{logsemColor}{RGB}{0,80,125}

\definecolor{numsemColor}{RGB}{204,100,0}

\newcommand{\qa}[1]{{\footnotesize\textit{{\textsf{#1}}}}}

\newcommand{\qaClose}[0]{\qa{close}}
\newcommand{\qaClosePred}[2]{\qaClose(#1, #2)}
\newcommand{\qaMed}[0]{\qa{medDist}}
\newcommand{\qaMedPred}[2]{\qaMed(#1, #2)}

\newcommand{\qaOnIncomingLaneOf}[0]{\qa{onIncomingLane}}
\newcommand{\qaOnIncomingLaneOfPred}[2]{\qaOnIncomingLaneOf(#1, #2)}

\newcommand{\qaHasValidHeading}[0]{\qa{validHeading}}
\newcommand{\qaHasValidHeadingPred}[1]{\qaHasValidHeading(#1)}

\newcommand{\qaOnDifferentLanes}[0]{\qa{differentLanes}}
\newcommand{\qaOnDifferentLanesPred}[2]{\qaOnDifferentLanes(#1, #2)}

\newcommand{\qaPotentiallyCollidesWith}[0]{\qa{PotentiallyCollidesWith}}
\newcommand{\qaPotentiallyCollidesWithPred}[2]{\qaPotentiallyCollidesWith(#1, #2)}

\newcommand{\qaCar}[0]{\qa{Car}}
\newcommand{\qaJunction}[0]{\qa{Junction}}

\newcommand{\qaManLeft}[0]{\qa{TurnLeft}}
\newcommand{\qaManLeftPred}[2]{\qaManLeft(#1, #2)}
\newcommand{\qaManStraight}[0]{\qa{GoStraight}}
\newcommand{\qaManStraightPred}[2]{\qaManStraight(#1, #2)}

\definecolor{blue0}{HTML}{47BCFF}
\definecolor{blue1}{HTML}{009BF5}
\definecolor{blue2}{HTML}{0067A3}
\definecolor{blue3}{HTML}{003452}
\definecolor{car1}{RGB}{128,125,123}  
\definecolor{car2}{RGB}{194,92,85}  
\definecolor{car3}{RGB}{75,119,157}  

\newcommand{\scenic}{\textsc{Scenic}}

\newcommand{\actorNM}[1]{\vec{\texttt{a}}_#1}
\newcommand{\actor}[1]{$\actorNM{#1}$}

\newcommand{\actorR}[0]{\textcolor{car2}{\actor{R}}}
\newcommand{\actorBl}[0]{\textcolor{car3}{\actor{B}}}

\newcommand{\justParam}[1]{\texttt{#1}}

\newcommand{\justParamBl}[1]{\textcolor{car3}{\justParam{#1}}}
\newcommand{\justParamR}[1]{\textcolor{car2}{\justParam{#1}}}

\newcommand{\param}[2]{\actorNM{#2}.\justParam{#1}}

\newcommand{\paramR}[1]{\textcolor{car2}{\param{#1}{R}}}
\newcommand{\paramBl}[1]{\textcolor{car3}{\param{#1}{B}}}

\newcommand{\actorFunNoMath}[1]{\texttt{o}_{#1}}

\newcommand{\actorFun}[1]{$\actorFunNoMath{#1}$}

\newcommand{\actorFunR}[0]{\textcolor{car2}{\actorFun{R}}}
\newcommand{\actorFunBl}[0]{\textcolor{car3}{\actorFun{B}}}

\newcommand{\juncFunNoMath}[1]{\texttt{j}_{#1}}
\newcommand{\juncFun}[1]{$\juncFunNoMath{#1}$}
\newcommand{\juncFunA}[0]{\textcolor{gray}{\juncFun{A}}}

\newmdenv[topline=false,bottomline=false,rightline=false,innertopmargin=1pt,innerbottommargin=1pt,innerrightmargin=0pt,innerleftmargin=0.65ex,skipabove=0.65ex,skipbelow=0.65ex,linewidth=0.75pt]{exlineEnv}
\newcounter{exline}
\newenvironment{exline}[1][]{\refstepcounter{exline}
\begin{exlineEnv}\noindent\textit{Example~\theexline: #1}\rmfamily}
{\end{exlineEnv}}

\newcolumntype{Z}{>{\centering\let\newline\\\arraybackslash\hspace{0pt}}X}

\newcommand{\rquestion}[1]{\textbf{\textsc{RQ}#1}}
\newcommand{\ranswer}[2]{\boxedVal{\rquestion{#1:}}{\emph{#2}}}
\newcommand{\contribution}[1]{\textbf{\textsc{C}#1}}

\newcommand{\thm}[1]{\textbf{\textsc{Theorem} #1}}
\newcommand{\theorembox}[2]{%
\vskip 0.4\baselineskip
\noindent
\begin{tabularx}{\linewidth}{X}
\thm{#1:} 
#2\\
\end{tabularx}
\vskip 0.4\baselineskip}

\newcommand{\boxedVal}[2]{%
\vskip 0.5\baselineskip
\noindent
\begin{tabularx}{\linewidth}{|X|}
\hline
#1 
#2\\\hline
\end{tabularx}}

\newcounter{enumi-saved}

\definecolor{completed}{RGB}{51,153,102}

\definecolor{inProgress}{RGB}{196,196,35}

\definecolor{notYetStarted}{RGB}{255,0,0}

\NewDocumentCommand{\rot}{O{45} O{1em} m}{\makebox[#2][l]{\rotatebox{#1}{#3}}}%
\newcommand{\halfcheck}{X\kern-1.1ex\raisebox{.7ex}{\rotatebox[origin=c]{125}{--}}}

\begin{document}

\title{Automated and Complete Generation of Traffic Scenarios at Road Junctions Using a Multi-level Danger Definition}

\author*[1,2]{\fnm{Aren~A.} \sur{Babikian}}
\email{babikian@cs.toronto.edu}

\author[3]{\fnm{Attila} \sur{Ficsor}}
\email{ficsor@mit.bme.hu}

\author[3]{\fnm{Oszk\'{a}r} \sur{Semer\'{a}th}}
\email{semerath@mit.bme.hu}

\author[1]{\fnm{Gunter} \sur{Mussbacher}}
\email{gunter.mussbacher@mcgill.ca}

\author[1,3,4]{\fnm{D\'{a}niel} \sur{Varr\'{o}}}
\email{daniel.varro@liu.se}
            
\affil[1]{\orgdiv{Department of Electrical and Computer Engineering}, \\\orgname{McGill University}, \orgaddress{\city{Montreal}, \country{Canada}}}

\affil[2]{\orgdiv{Department of Computer Science}, \orgname{University of Toronto}, \\\orgaddress{\city{Toronto}, \country{Canada}}}

\affil[3]{\orgdiv{Department of Artificial Intelligence and Systems Engineering}, \orgname{Budapest University of Technology and Economics}, \orgaddress{\city{Budapest}, \country{Hungary}}}

\affil[4]{\orgdiv{Department of Computer and Information Science}, \orgname{Link\"{o}ping University}, \orgaddress{\city{Link\"{o}ping}, \country{Sweden}}}


\abstract{

To ensure their safe use, autonomous vehicles (AVs) must meet rigorous certification criteria that involve executing maneuvers safely within (arbitrary) scenarios where other actors perform their intended maneuvers.
For that purpose, existing scenario generation approaches optimize search to derive scenarios with high probability of dangerous situations. 
In this paper, we hypothesize that at road junctions, potential danger predominantly arises from overlapping paths of individual actors carrying out their designated high-level maneuvers.
As a step towards AV certification, we propose an approach to derive a complete set of (potentially dangerous) abstract scenarios at any given road junction, i.e. all permutations of overlapping abstract paths assigned to actors (including the AV) for a given set of possible abstract paths.
From these abstract scenarios, we derive exact paths that actors must follow 
to guide simulation-based testing towards potential collisions.   
We conduct extensive experiments to evaluate the behavior of a state-of-the-art learning-based AV controller on scenarios generated over two realistic road junctions with increasing number of external actors.
Results show that the AV-under-test is involved in increasing percentages of unsafe behaviors in simulation, which vary according to functional- and logical-level scenario properties. 
}




\maketitle


\section{Introduction}

\textbf{Motivation:}
The increasing popularity of autonomous vehicles (AVs) has resulted in a rising interest in their safety assurance.
As prescribed in safety standards for road vehicles \citep{ISO26262FunctionalSatefy,ISO21448SOTIFRoadVehicles},
rigorous certification criteria 
must be met by AVs to ensure their safe, widespread use from a societal perspective.
In this context, existing research \citep{Czarnecki2018Taxonomy,Majzik2019TowardsSystemLevel} particularly emphasizes the importance of adhering to \textit{system-level} safety requirements.

Early research suggests neither upfront design-time verification nor on-road monitoring
is practically feasable \citep{Kalra2016,Koopman2016,Helle2016} for AV certification.
As such, existing safety assurance approaches \citep{Abdessalem2018TestingFeatureInteractionsBriandNsgaForConcreteScenes,Babikian2021dReal} adopt the scenario-based testing paradigm: they test AVs by (1) generating challenging traffic scenarios, (2) executing them in simulation, and (3) evaluating the system-level safety of the AV-under-test.


\textbf{Problem statement:}
As AVs interact with a complex and dynamically changing environment, they may encounter a potentially infinite number of concrete scenarios, which is infeasible for certification.
As such, rigorous safety assurance needs to consider a \textit{(A) complete set of (B) practically relevant abstract scenarios} \citep{babikianConcretizationAbstractTraffic2024}.

To adequately measure the \textit{completeness of a test suite for certification} purposes, existing research suggests using coverage criteria on a high level of abstraction (e.g. situation coverage \citep{Alexander2015SituationCoverage}).
In particular, such criteria are measured over abstract scenario representations \citep{Ulbrich2015Defining,Majzik2019TowardsSystemLevel} to reduce the infinite search space of possible scenarios.



To generate \textit{practically relevant} scenarios, many test scenario generation approaches
\citep{Abdessalem2018TestingVisionBased,Abdessalem2018TestingFeatureInteractionsBriandNsgaForConcreteScenes,Haq2023Reinforcement,Zhong2021Fuzzing,calo2020GeneratingAvoidableCollision,Wu2021} use scenario-specific danger criteria as guiding metrics to derive concrete refinements of the same abstract scenario. 
While such approaches often expose bugs in AV behavior, they fail to answer a key certification question: have we tested enough scenarios?
The use of concrete guidance metrics limits the abstract coverage of a generated test suite.
To obtain a test suite with high abstract coverage (i.e. to satisfy broader safety assurance criteria), an abstract guidance metric is required.

Furthermore, such test scenario generation approaches are tailored to concretize a single abstract scenario.
For instance, \citet{calo2020GeneratingAvoidableCollision} generate colliding scenarios at a 4-way road junction where “the [AV-under-test proceeds] on its lane and two cars [cross] the main road from left to right”.
Generating a test suite with high abstract coverage requires deriving concrete scenarios from a variety of abstract scenarios, either given as input or derived a priori.





\textbf{Contributions:}
Our paper presents an approach to automatically derive a set of dangerous concrete traffic scenarios with abstract coverage guarantees over the set of possible maneuvers at a road junction given as input.
We focus on single-maneuver scenarios at road junctions, where the set of possible (simple) maneuvers (e.g. left turns, right turns, straight drives through the junction) is varied and well-defined.
Maneuver sequences are outside the scope of this paper.

\begin{itemize}

    \item[\contribution{1}:] We synthesize\textit{ a complete set of avoidable, collision-inducing logical scenarios}
    that cover 
    any combination of high-level maneuvers of the AV and other vehicles at a road junction.
    
    
    \item[\contribution{2}:] We lift a concrete definition of danger (i.e. potential existence of a collision) to \textit{a higher level of abstraction} and use it as a heuristic to derive dangerous logical-level scenarios.
    
    
    

    \item[\contribution{3}:] We refine logical scenarios into \textit{collision-inducing exact paths as concrete scenarios amenable to simulation} for various actors parameterized by a high-level maneuver and a concrete speed/acceleration profile given as input.
    
    \item[\contribution{4}:] We \textit{extensively evaluate} our approach \textit{in simulation involving a state-of-the-art AV controller} over two 
    road junctions with diverse characteristics to analyse dangerous situations at various levels of abstraction.
\end{itemize}


The rest of the paper is structured as follows:
\autoref{sec:preliminaries} summarizes core concepts related to traffic scenarios in AV safety assurance.
\autoref{sec: overview} provides an overview of our approach. 
\autoref{sec:Logical} presents our logical scenario generation approach.
\autoref{sec:Concrete} details the generation of concrete scenarios from logical scenarios.
\autoref{sec:evaluation} provides evaluation results of our approach. 
\autoref{sec:relWork} discusses related work.
Finally, \autoref{sec:conclusion} concludes the paper.

\section{Preliminaries}
\label{sec:preliminaries}

\subsection{Traffic scenes and scenarios}
\label{sec:pre-scene}

\textit{Traffic scene} is defined by \citet{Ulbrich2015Defining} as a snapshot of the environment, including the \textit{scenery} and \textit{dynamic elements}, as well as the \textit{relations} between those entities.

The \textit{scenery} is comprised of the lane network, stationary components such as traffic lights and curbs, vertical elevation of roads and environmental conditions.
\textit{Dynamic elements} (or \textit{actors}) include the various vehicles and pedestrians
involved in a scene. This includes the \textit{ego} actor under-test and, optionally, \textit{external} actors (e.g. other vehicles, pedestrians). A scene may contain information about the state (e.g. position and speed) and attributes (e.g. vehicle color, whether a car door is open) of actors.
 \textit{Relations} are defined between scenery elements and/or actors. For example, 
two vehicles may be \textit{far from each other}, or a vehicle may be placed \textit{on} a specific lane. 

A sequence of consecutive traffic scenes together with related temporal developments constitutes a \textit{scenario}, defined by an \textit{initial scene}, followed by a sequence of \textit{actions and events} performed by the actors according to their individual \textit{goals}.
\textit{Actions and events} often refer to high-level, qualitative maneuvers (e.g. a left turn at a junction, a lane change, an overtaking maneuver), while \textit{goals} may be transient (e.g. reaching a certain area on a map) or permanent (e.g. driving in a safe manner).

Traffic scenarios are commonly used for the simulation-based testing of AVs.
In this context, each actor initiates a sequence of high-level, actor-specific maneuvers.
For \textit{external actors}, such maneuvers are enforced: if a specific path and speed is assigned to an external actor, it \textit{must} follow the assigned path at the given  speed without considering any external factors, e.g. road safety.
However, the \textit{ego actor} is assigned a particular \textit{goal} (which is often times to adhere to traffic safety requirements \citep{Fremont2019ScenicLanguage,Abdessalem2018TestingVisionBased,Haq2022,Riccio2020}).
Additionally, its assigned actions (e.g. a particular sequence of maneuvers) are interpreted as (adaptable) \textit{instructions}: the ego actor must follow the assigned actions but may (partially) deviate from them only if it is necessary for the ego actor to reach its (safety) goals.
For instance, the ego vehicle may slow down during a left turn if it detects an oncoming vehicle that might cause a collision.


\subsection{Levels of abstraction in traffic scenarios}
\label{sec:pre-levels}
In this paper, we consider traffic scenarios on three levels of abstraction \citep{Menzel2018ScenariosForDevelopment}, as prescribed by existing guidelines to adequately describe traffic scenario for AV simulation \citep{Majzik2019TowardsSystemLevel}.
\begin{enumerate}
    \item \emph{Functional scenarios} include abstract (qualitative) constraints for    
    geospatial concepts (e.g. two vehicles are \emph{close to} each other), 
    causal concepts (e.g. a vehicle stopped \emph{because} it encountered a red light) and temporal concepts (e.g. two vehicles will \emph{eventually} collide).
    \item \emph{Logical scenarios} refine the abstract constraints of functional scenarios into constraints over parameter ranges or intervals, optionally specified by probability distributions. For example, geospatial functional constraints may be refined to areas on a map, and temporal functional constraints may be refined to time intervals.
    \item \emph{Concrete scenarios} refine the parameter ranges/intervals defined in a logical scenario to exact numeric values, e.g., for the position coordinates of actors. similarly, event executions are refined to exact times and durations.
\end{enumerate}


\begin{figure}[htp]
\noindent
\captionsetup{justification=centering}
\footnotesize
\begin{tabular}{
  |m{\dimexpr.50\linewidth-2\tabcolsep-1.3333\arrayrulewidth}
  |m{\dimexpr.50\linewidth-2\tabcolsep-1.3333\arrayrulewidth}|
  }
\hline

\multicolumn{2}{|c|}{Functional Scenario}\\
\hdashline
\multicolumn{2}{|c|}{
\makecell[l]{
\actorFunR~: \qaCar. \actorFunBl~: \qaCar. \juncFunA~: \qaJunction.\\
\textcolor{commentcolor}{\# Initial Scene Constraints} \\
\qaOnIncomingLaneOfPred{\actorFunR}{\juncFunA}. \qaHasValidHeadingPred{\actorFunR}. \qaClosePred{\juncFunA}{\actorFunR}.\\
\qaOnIncomingLaneOfPred{\actorFunBl}{\juncFunA}. \qaHasValidHeadingPred{\actorFunBl}. \qaMedPred{\juncFunA}{\actorFunBl}.\\
\qaOnDifferentLanesPred{\actorFunR}{\actorFunBl}.\\
\textcolor{commentcolor}{\# Behavioral Constraints} \\
\qaManLeftPred{\actorFunR}{\juncFunA}.\qaManStraightPred{\actorFunBl}{\juncFunA}.\\
\qaPotentiallyCollidesWithPred{\actorFunR}{\actorFunBl}.\\
}
}\\

\hline

Logical Scenario
&
Concrete Scenario \\
\hdashline

\makecell[l]{
\actorFunR $\mapsto$ \actorR = $\langle \justParamR{x}, \justParamR{y}, \justParamR{h}, \justParamR{spd}\rangle$; \\
\actorFunBl $\mapsto$ \actorBl = $\langle \justParamBl{x}, \justParamBl{y}, \justParamBl{h}, \justParamBl{spd}\rangle$ \\
\textcolor{commentcolor}{\# Initial Conditions} \\
$ \paramR{spd} \in [10..30]$; 
$ \paramBl{spd} \in [30..50]$\\
\textcolor{commentcolor}{\# Expected Path Regions} \\
\includegraphics[valign=c,margin=0pt 1ex 0pt 1ex,width=\linewidth]{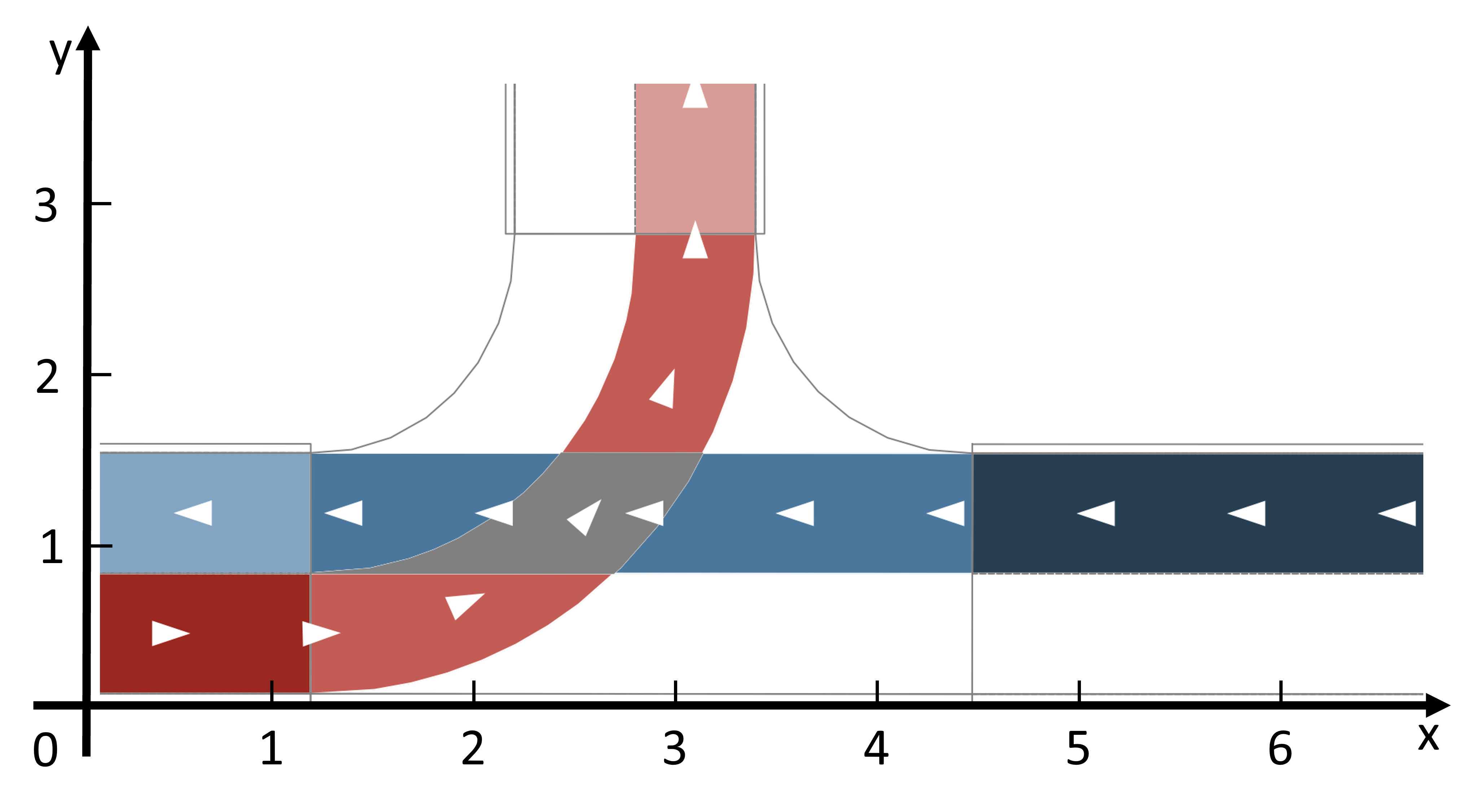}\\
}

&

\makecell[l]{
\textcolor{commentcolor}{\# Initial Conditions} \\
\actorR = $\langle \justParamR{x}{=}0.6, \justParamR{y}{=}0.5, \justParamR{h}{=}0, \justParamR{S}{=}20\rangle$ \\
\actorBl = $\langle \justParamBl{x}{=}6.3, \justParamBl{y}{=}1.2, \justParamBl{h}{=}\pi, \justParamBl{S}{=}35\rangle$ \\
\\
\textcolor{commentcolor}{\# Expected Exact Paths} \\
\includegraphics[valign=c,margin=0pt 1ex 0pt 1ex,width=\linewidth]{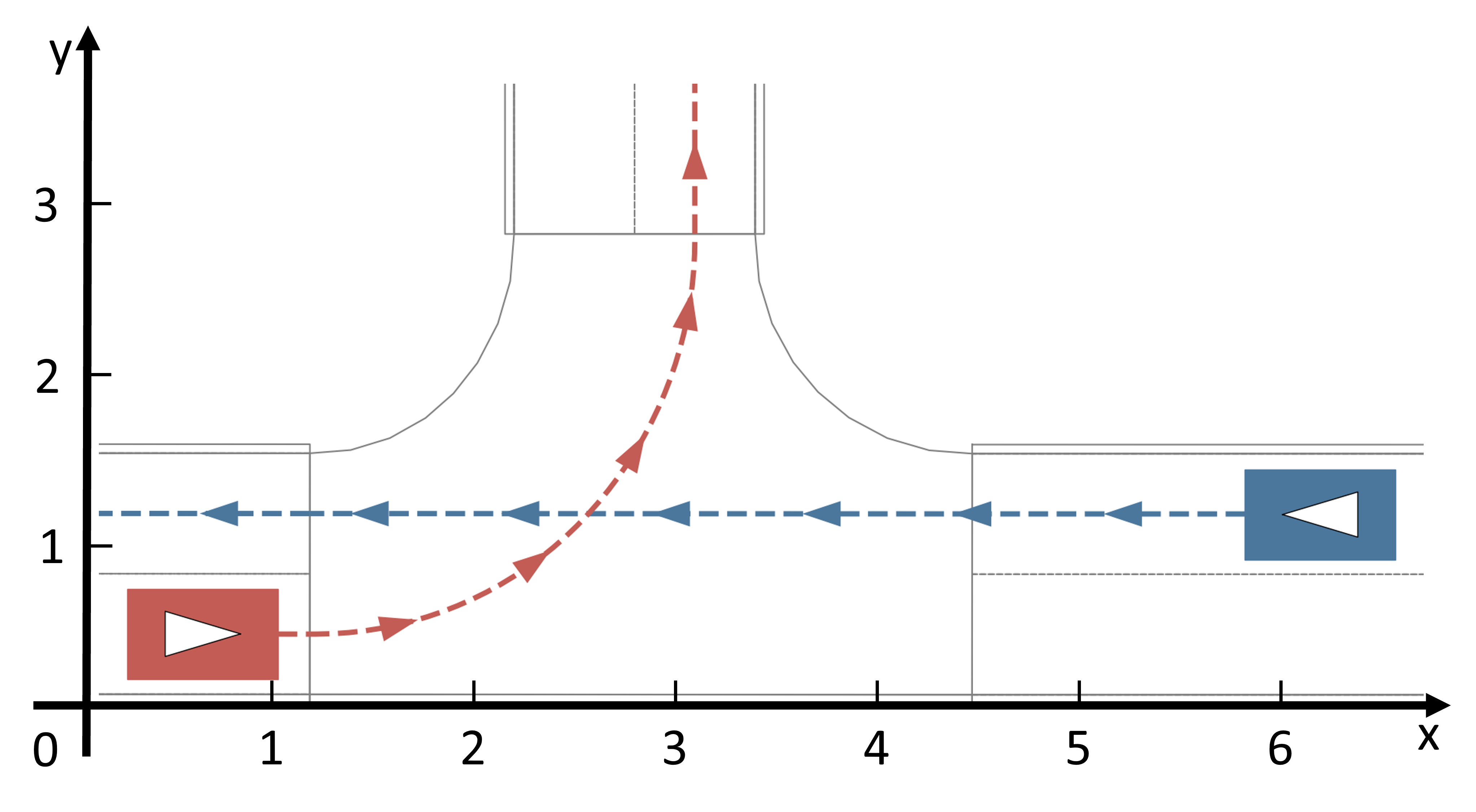}\\
}\\
\hline
\end{tabular}
\caption{A traffic scenario represented at various levels of abstraction}
\label{fig:scenarioAbsLevels}
\end{figure}

\begin{exline}
\autoref{fig:scenarioAbsLevels} describes (at various levels of abstraction) a traffic scenario representing a practically relevant test case that involves two (dynamic) actors (\actorFunR~and \actorFunBl) at a road junction \juncFunA.
\begin{enumerate}
    \item At the \textbf{functional} level, the scenario is described using high-level constraints:
    \actorFunR~ and \actorFunBl~are placed on different incoming lanes of \juncFunA~ and have valid headings.
    They are initially placed at a certain distance from \juncFunA~with a specific maneuver 
    to perform while inside \juncFunA.
    Additionally, as a testing objective (i.e. to evaluate actor behavior), both vehicles are expected to potentially collide with each other.
    Note that the functional scenario is presented in a formal syntax \citep{babikianConcretizationAbstractTraffic2024} based on partial-model semantics.
    \item The \textbf{logical} scenario assigns colliding path regions (lane sequences) to the actors.
    The initial regions (depicted in a darker shade) and initial speed intervals are also included.
    \item The \textbf{concrete} scenario determines the exact starting position, heading and speed for each vehicle. Additionally, it decides upon the exact path for each actor, depicted in their corresponding color. These exact parameters are selected such that a collision would occur between both actors, as required by the functional scenario.
\end{enumerate}
\end{exline}

\begin{figure}[htp]
    \begin{center}
      \includegraphics[width=.8\linewidth]{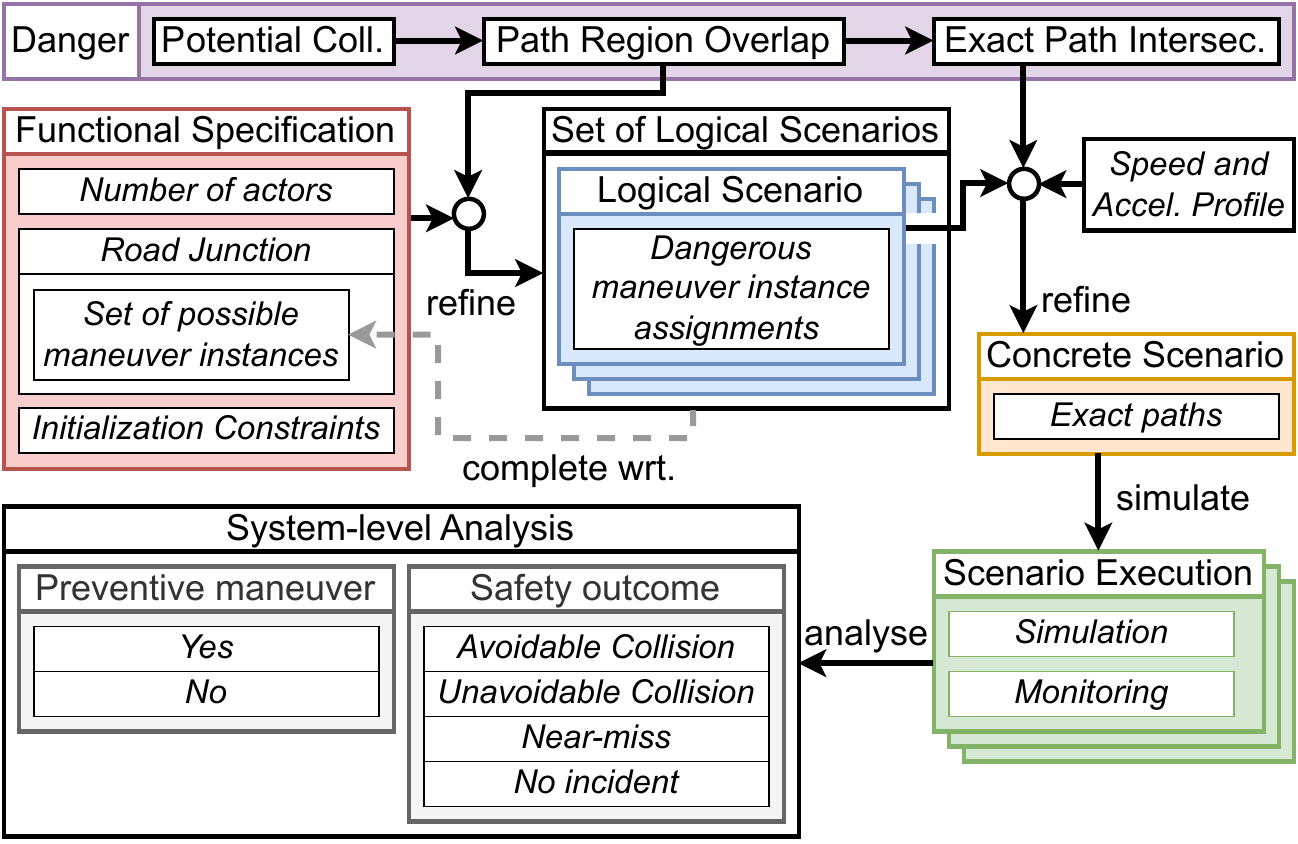}
      \caption{Overview of our scenario generation approach}
      \label{fig:overview}
    \end{center}
\end{figure}

\section{Overview of our approach}
\label{sec: overview}




\textbf{Functional-to-logical refinement:}
Our approach (depicted in \autoref{fig:overview}) for generating dangerous scenarios takes as input a \textit{functional-level scenario specification} which is defined by (1) the number of actors involved in the scenario, (2) the road junction under test and (3) constraints over the initial scene of the scenario.
The road junction includes a set of all possible maneuver instances, which, in our case, corresponds to the set of all possible left turns, right turns and straight drive maneuvers at the junction.
Initialization constraints require (a) to initially place vehicles on an incoming road of the junction and (b) to assign them a heading aligned with the intended road map orientation of its initial lane.


Guided by our definition of danger, the functional-level scenario 
is then refined into \textit{a set of logical scenarios}, as detailed in \autoref{sec:Logical}.
At the logical level, scenarios are represented as a mapping between actors and their designated maneuver instance (among those listed at the functional level) such that danger is enforced.

As a novel contribution (\contribution{1}), our approach generates the complete set of collision-inducing logical scenarios over the space of actor-specific maneuvers 
given as input.
To the best of our knowledge, we are the first to integrate such a measure of \textit{abstract} coverage into a concrete scenario generation approach, which makes our approach particularly beneficial for AV certification.
Furthermore, our approach uses an abstract (logical-level) danger definition to guide scenario generation that is applicable to a variety of functional scenario specifications containing any number of actors at any road junction regardless of its geometry (contribution \contribution{2}).


\textbf{Defining danger:}
At the functional level, we consider a scenario to be dangerous if it places the ego actor in a \textit{potentially collision-inducing situation} with an external actor.
An increasing number of external actors potentially colliding with the ego actor results in an increasing level of danger in the context of the scenario.

We refine this definition of danger to the logical level, where a collision-inducing situation between two actors is represented as the \textit{overlap of the path regions} associated to the maneuver of each actor. 
Overlapping path regions do not ensure the existence of a collision, which depends on concrete-level properties (e.g. designated speed and acceleration profiles) and simulated behavior (e.g. possible preventive maneuvers performed by an intelligent AV controller).
However, with the assumption that the ego actor does not deviate outside of its designated path region (unless it encounters a dangerous situation), the overlapping of path regions is a \textit{necessary condition} for a collision to occur.
Such an assumption is reasonable for real AV controllers, considering that they are often \textit{trained} to follow a designated sequence of points on a road map.

At the concrete level, a dangerous scenario requires the exact paths assigned to a pair of actors to \textit{intersect}.
Additionally, the concrete speeds and accelerations of the concerned actors are considered to ensure that they reach the intersecting point at the same time
(hence causing a potential collision, unless the involved ego actor initiates collision-avoiding actions).




\textbf{Logical-to-concrete refinement:}
As detailed in \autoref{sec:Concrete}, each logical scenario is then refined into a \textit{concrete scenario} where actors are assigned an exact path to follow.
The derived concrete scenario takes into consideration the concrete-level danger definition (i.e. exact path assignments between the ego actor and each external actor must intersect) as well as a speed and acceleration profile assigned to each actor (given as input).


As a novel contribution (\contribution{3}), our approach derives dangerous concrete scenarios amenable to simulation by refining functional scenarios with actors parameterized by high-level maneuvers.
This conceptually extends existing scenario generation approaches \citep{Abdessalem2018TestingVisionBased,Abdessalem2018TestingFeatureInteractionsBriandNsgaForConcreteScenes,Haq2023Reinforcement,Zhong2021Fuzzing,calo2020GeneratingAvoidableCollision,Wu2021}, which are tailored to concretize a single functional scenario without abstract completeness guarantees. 


\textbf{Scenario execution and analysis:}
As part of contribution \contribution{4}, each concrete scenario is then executed in \textit{simulation} and actor behavior is \textit{monitored}.
In \autoref{sec:evaluation}, we perform system-level analysis of the observed ego actor behavior.
We evaluate the \textit{safety outcome} of each simulation run: a run may result in a collision, a near-miss situation or in no incident.
A collision may be avoidable or unavoidable, as determined by exact in-simulation behavior.
As part of our system-level analysis, we also measure whether the ego actor performs preventive maneuvers to avoid potential collisions.
 
\textbf{Generalization:}
In this paper, the scenario generation process is illustrated in the context of road junctions due to their well-defined set of possible (simple) maneuvers (e.g. left/right turns, straight drives through the junction).
Nevertheless, the conceptual framework we propose can handle any road map component that employs the abstractions discussed in this paper, i.e. (logical) path regions and (concrete) exact paths.
The same is applicable when extending our approach to support different definitions of danger.

\section{Logical-level path regions}
\label{sec:Logical}

\label{sec:logical-desciription}

First, we propose an algorithm that refines an input functional scenario, and provides a set of logical scenarios.
At the logical level, each actor is assigned a lane-based \emph{path region} on the road map which \textit{over-approximates} all possible concrete paths of the actor's functional-level maneuver (while satisfying functional scenario requirements).
A path region represents the entirety (i.e. all points forming the path) of all possible concrete paths that may correspond to the given maneuver of the actor.
Such path regions correspond to constrained, 2-dimensional curves over the space of the road map, which is consistent with the definition proposed in \autoref{sec:pre-levels}.

In this paper, we generate scenarios where actor behavior is characterised by in-junction assigned maneuvers.
However, for adequate scenario execution, we must also consider actor behavior \textit{leading up to} and \textit{following} their in-junction maneuver.
As a result, logical-level path regions assigned to actors also include connected portions outside the road junction.

\newcommand{\jut}[0]{\juncFun{ut}}



\begin{exline}
The bottom-left cell of \autoref{fig:scenarioAbsLevels} shows two logical-level path regions assigned to two actors of a generated scenario.
Each actor must (1, dark) drive towards a road junction while following a lane (i.e. without performing a lane change), then (2, regular shade) perform a specified maneuver (\qa{turnLeft} and \qa{goStraight} for the red and blue actors, respectively) at the junction to finally (3, light) continue driving in the target lane until a termination criteria is reached.


\end{exline}

\newcommand{\maninst}[0]{MI}
\textbf{Deriving logical scenarios:}
Our approach takes a fixed number of actors and a road junction as input, and derives from it a set of \textit{dangerous logical scenarios} (i.e. collision-inducing scenarios, with overlapping path regions).
The derived set of logical scenarios is complete wrt. to the set of possible maneuver instances at the junction-under-test given as input.
In this context, a \textit{maneuver instance} (\maninst{}) corresponds to a specific path region through the junction-under-test, with specific start and end lanes and an associated maneuver type (e.g. a right turn).

\begin{figure}[htp]
     \centering
        \begin{subfigure}[b]{0.69\textwidth}
            \centering
            \begin{tabularx}{\linewidth}{|X|X|X|}\hline
                \includegraphics[valign=c,width=\linewidth-0.5\tabcolsep]{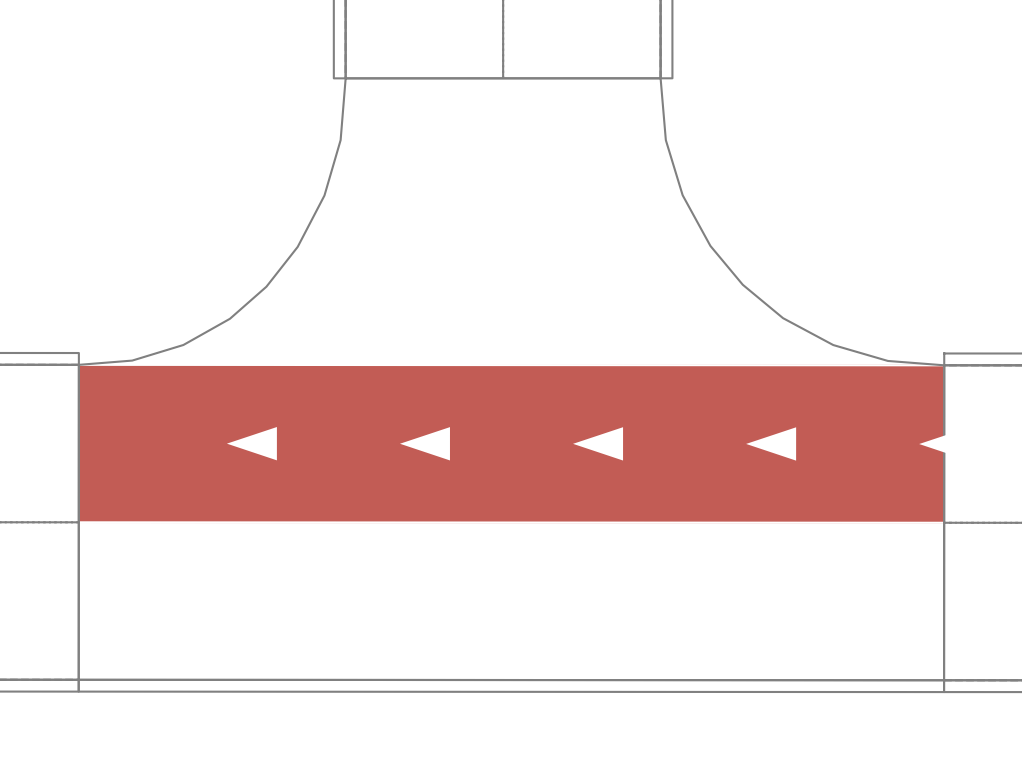} &
                \includegraphics[valign=c,width=\linewidth-0.5\tabcolsep]{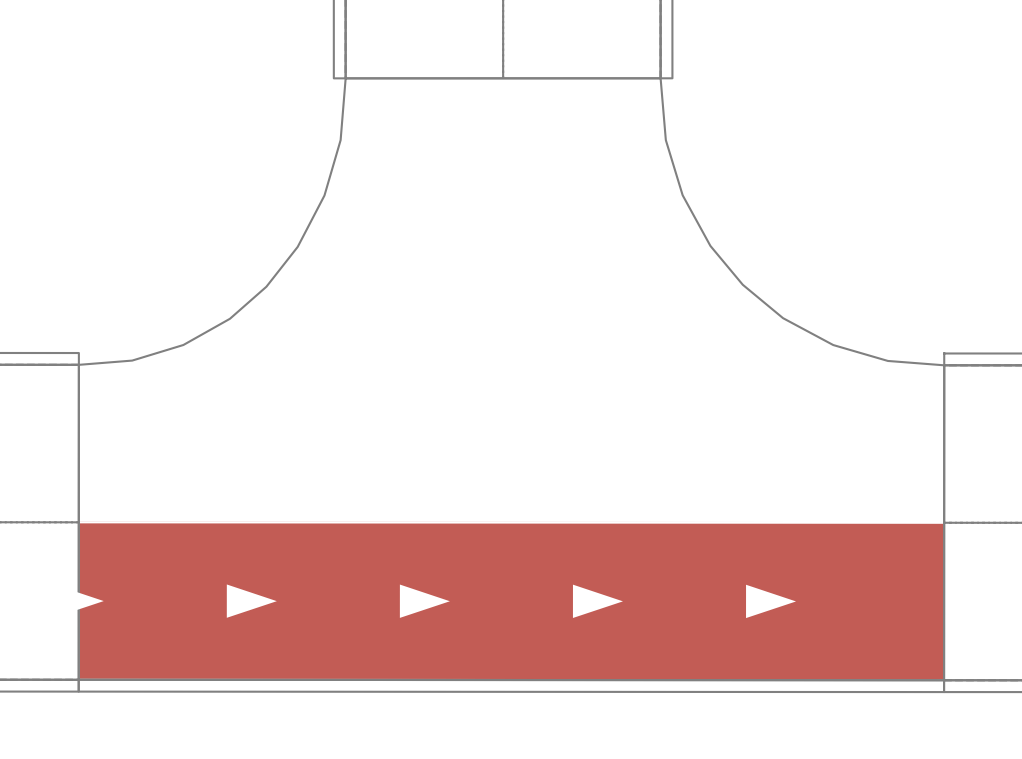} &
                \includegraphics[valign=c,width=\linewidth-0.5\tabcolsep]{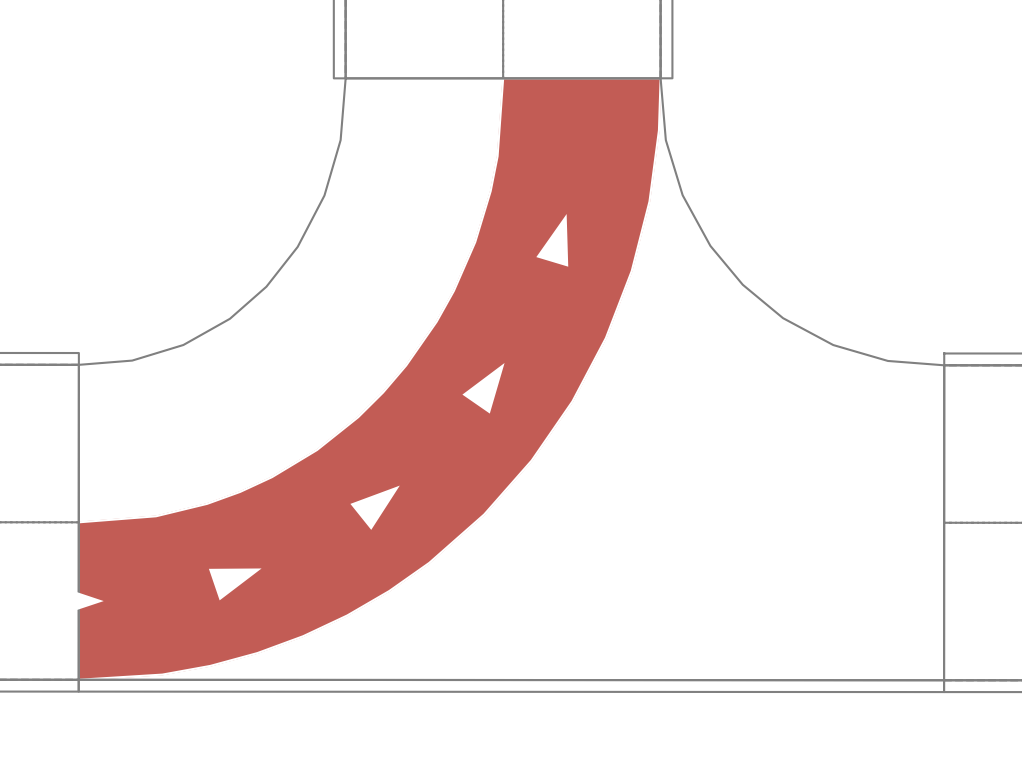} \\
                \includegraphics[valign=c,width=\linewidth-0.5\tabcolsep]{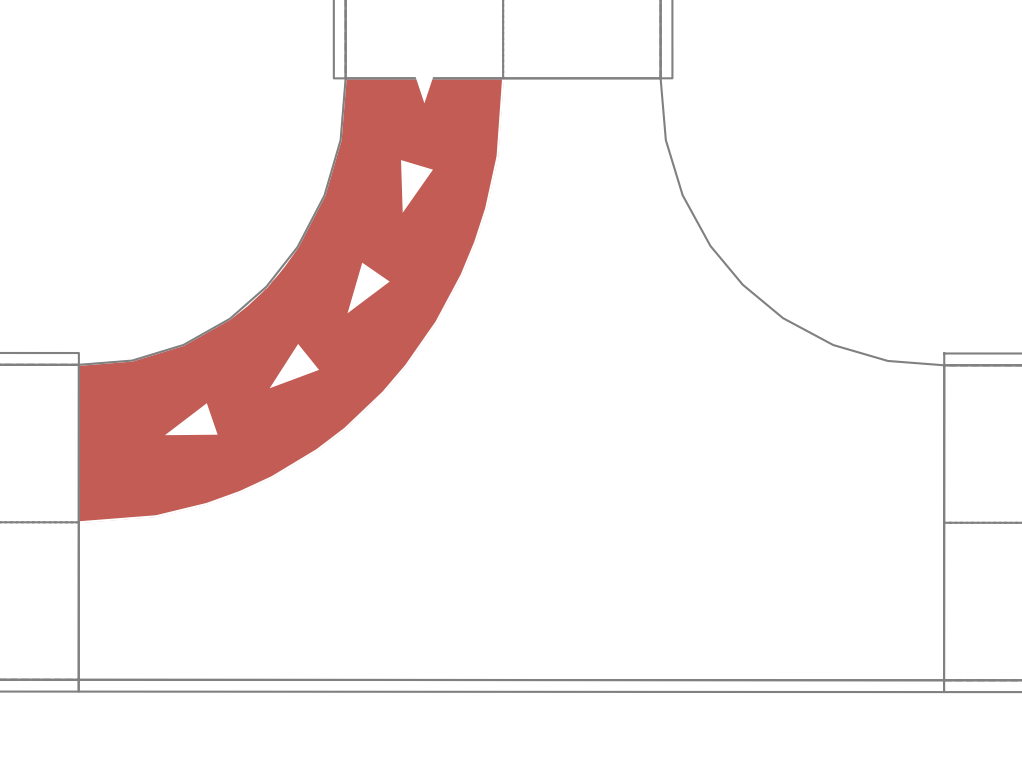} &
                \includegraphics[valign=c,width=\linewidth-0.5\tabcolsep]{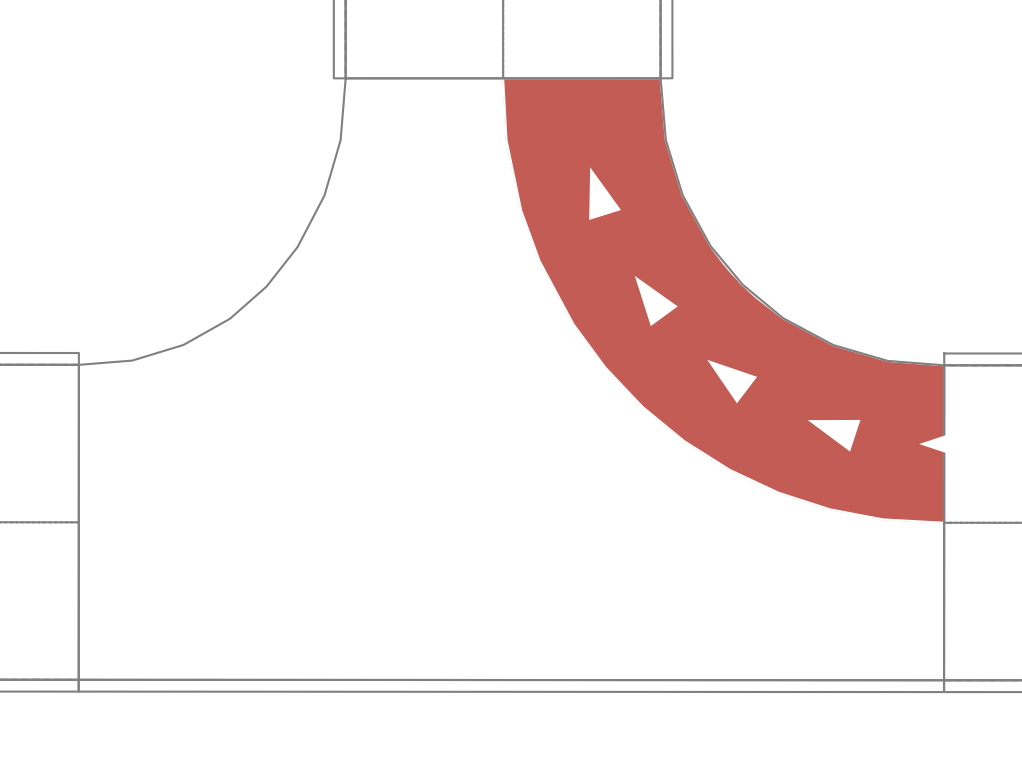} &
                \includegraphics[valign=c,width=\linewidth-0.5\tabcolsep]{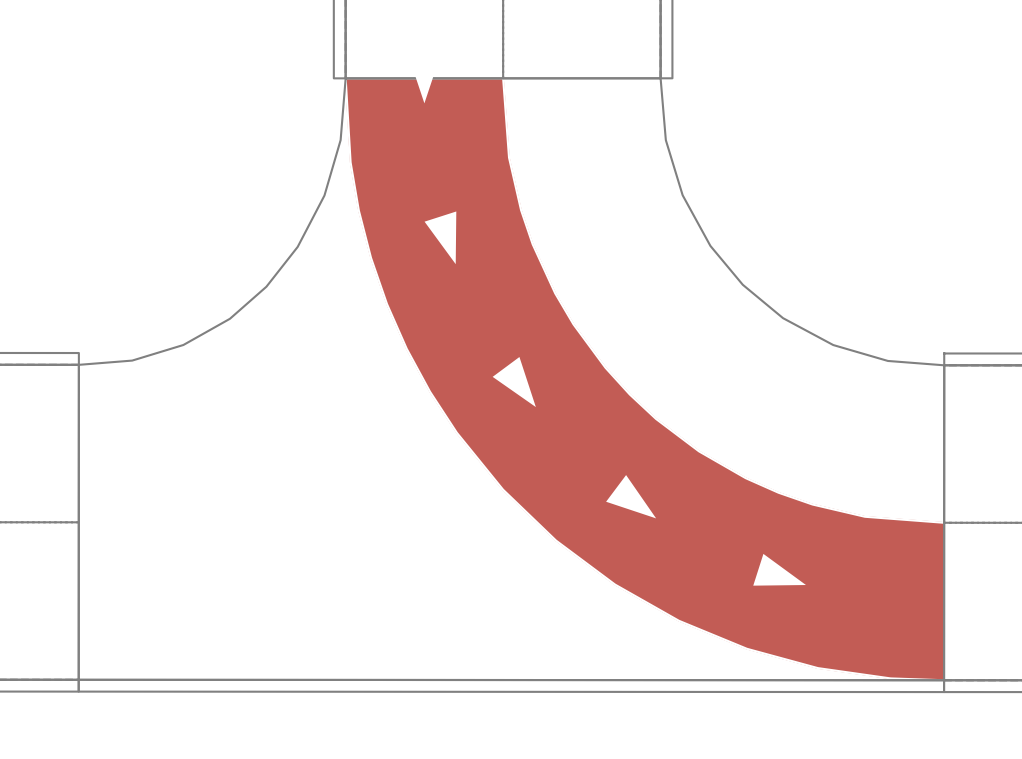} \\
                \hline
            \end{tabularx}
            \caption{Path regions associated to possible maneuver instances \\ \phantom{placeholder}}    
            \label{fig:regions-1-actor}
        \end{subfigure}
        \begin{subfigure}[b]{0.23\textwidth}
            \centering
            \begin{tabularx}{\linewidth}{|X|}\hline
                \includegraphics[valign=c,width=\linewidth-0.5\tabcolsep]{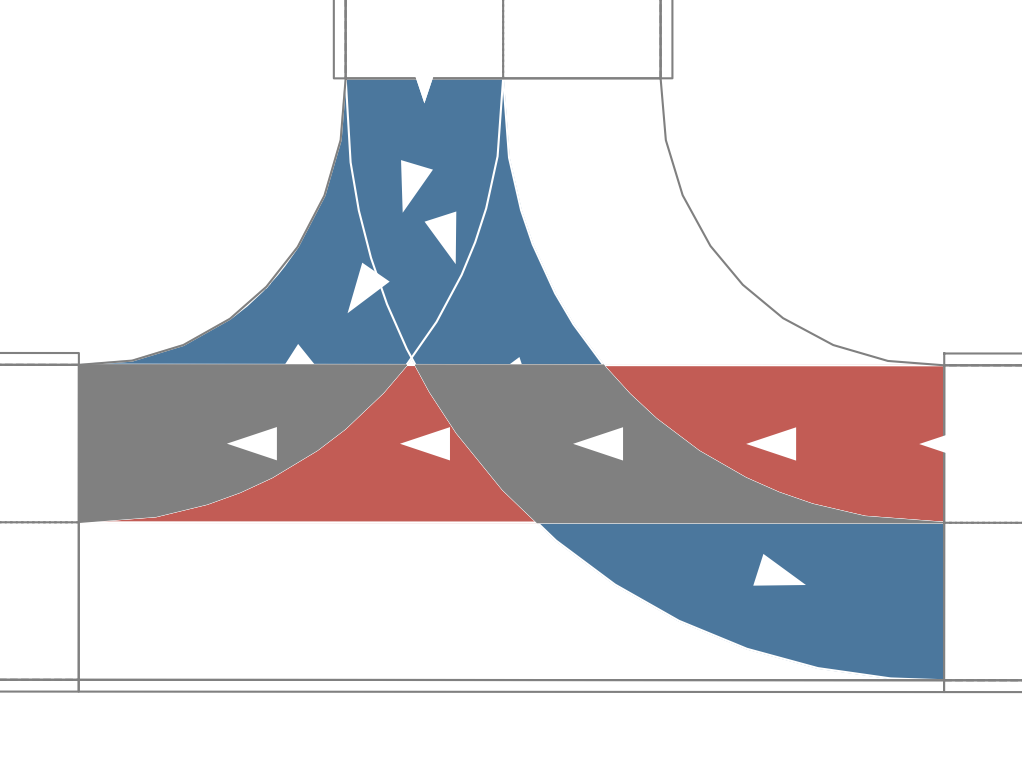} \\
                \includegraphics[valign=c,width=\linewidth-0.5\tabcolsep]{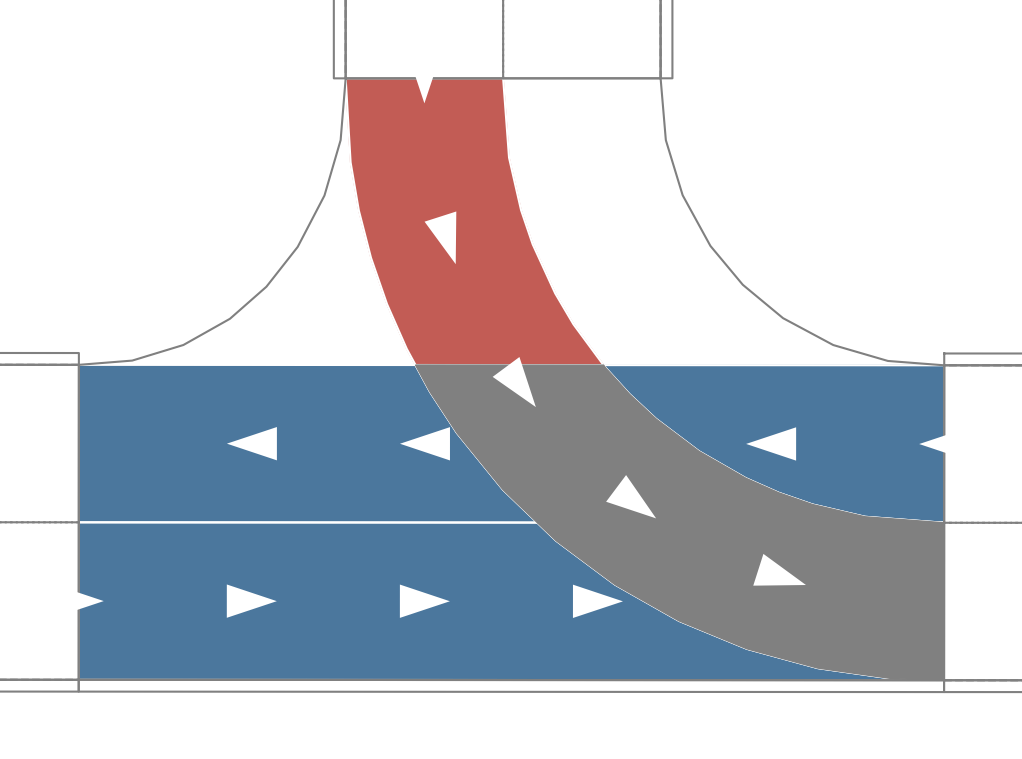} \\
                \hline
            \end{tabularx}
            \caption[]{Collision inducing 3-actor scenarios}    
            \label{fig:regions-3-actor-valid}
        \end{subfigure}
        \begin{subfigure}[b]{0.69\textwidth}
            \centering
            \begin{tabularx}{\linewidth}{|X|X|X|}\hline
                \includegraphics[valign=c,width=\linewidth-0.5\tabcolsep]{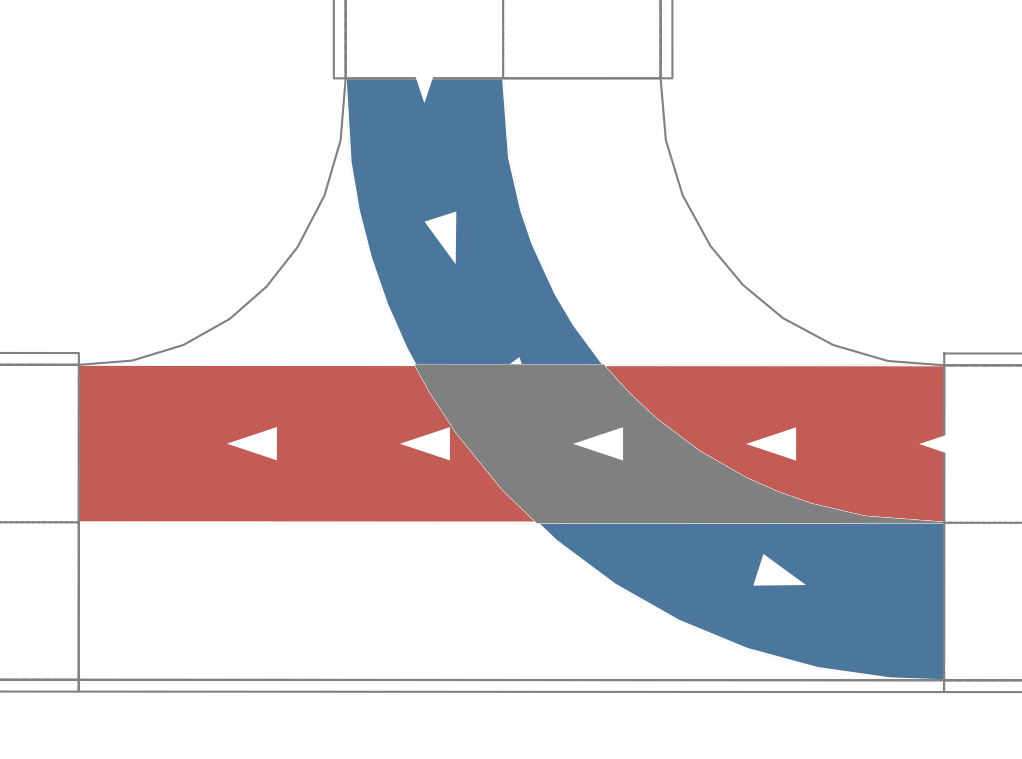} &
                \includegraphics[valign=c,width=\linewidth-0.5\tabcolsep]{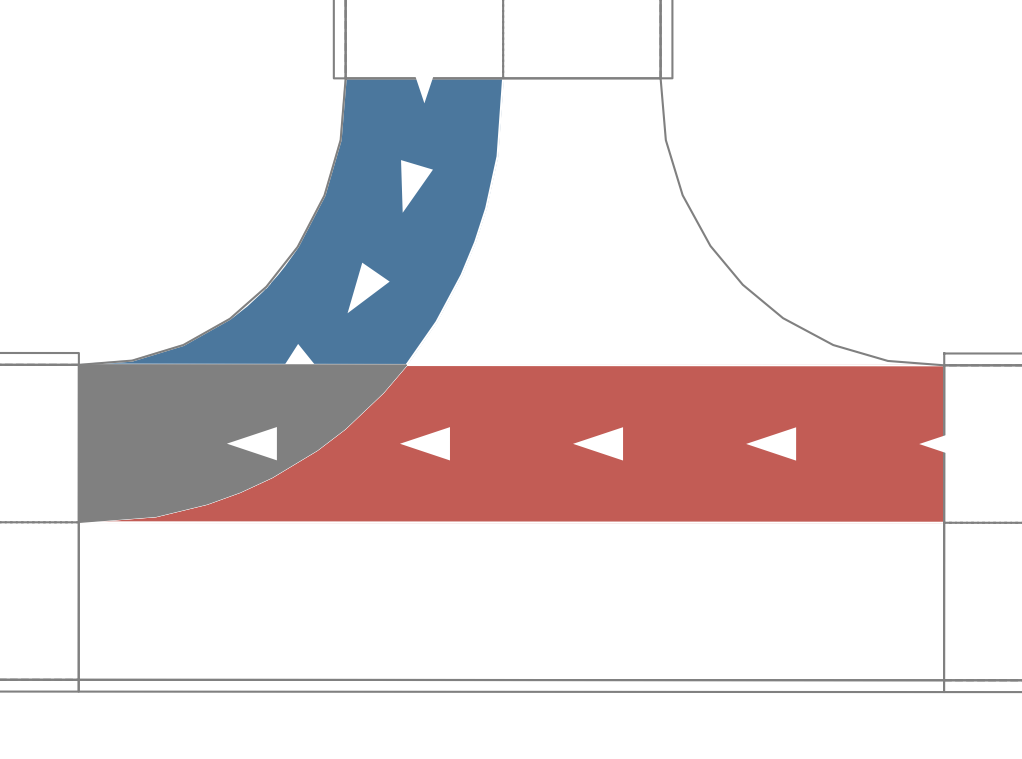} &
                \includegraphics[valign=c,width=\linewidth-0.5\tabcolsep]{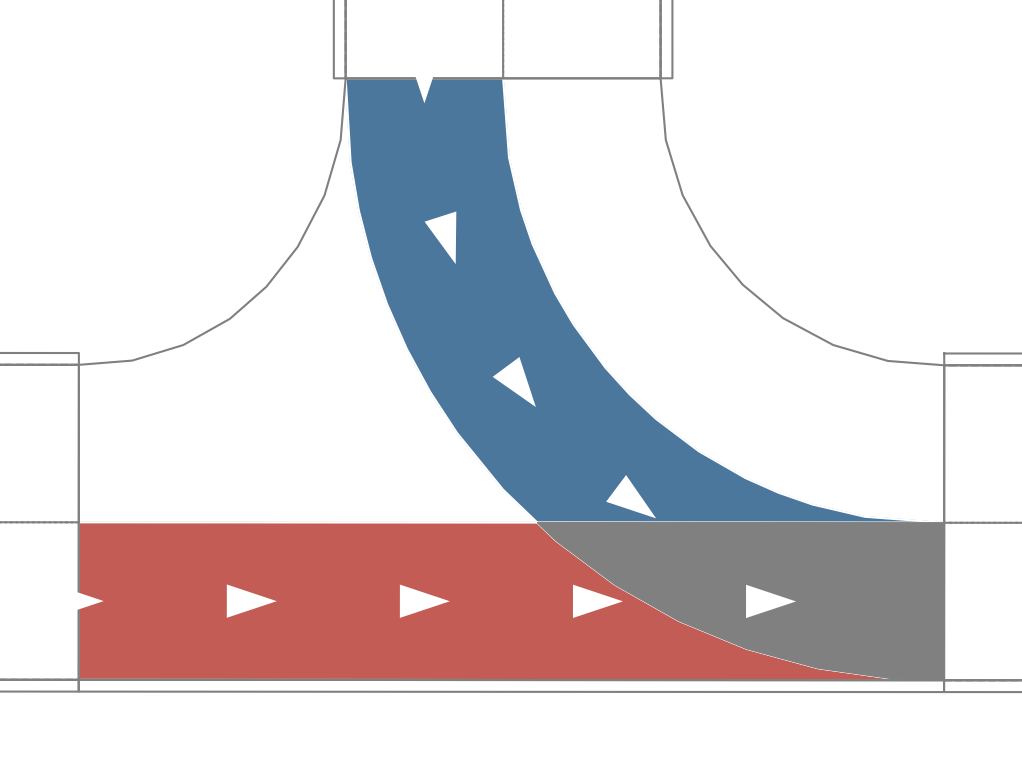} \\
                \includegraphics[valign=c,width=\linewidth-0.5\tabcolsep]{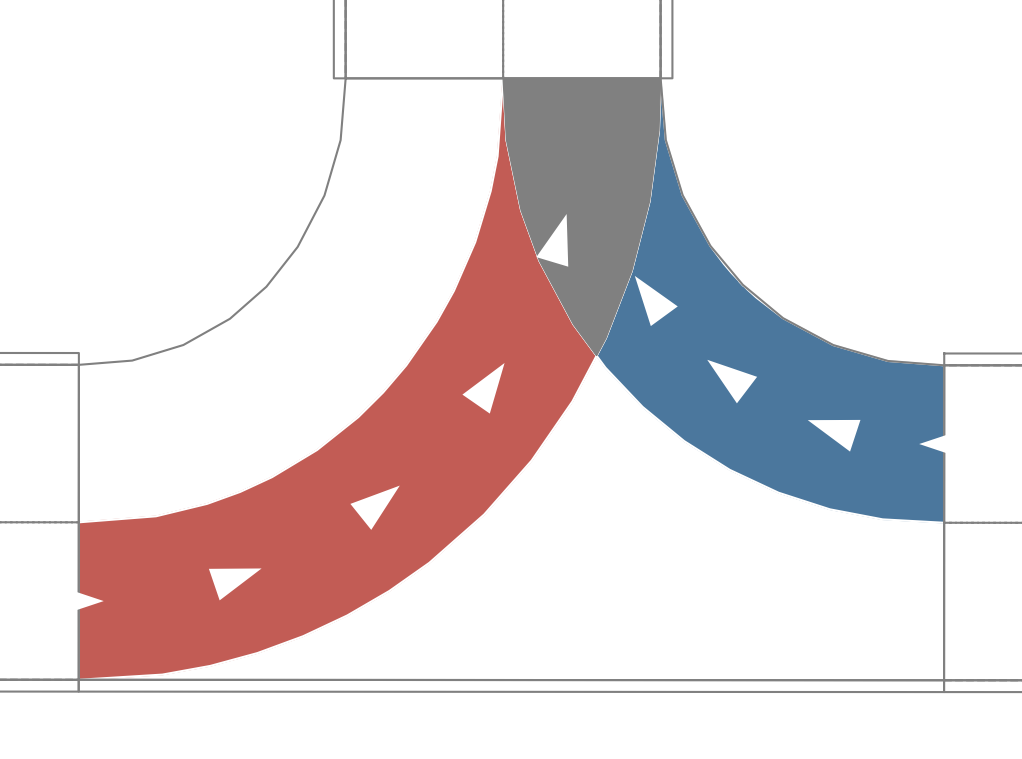} &
                \includegraphics[valign=c,width=\linewidth-0.5\tabcolsep]{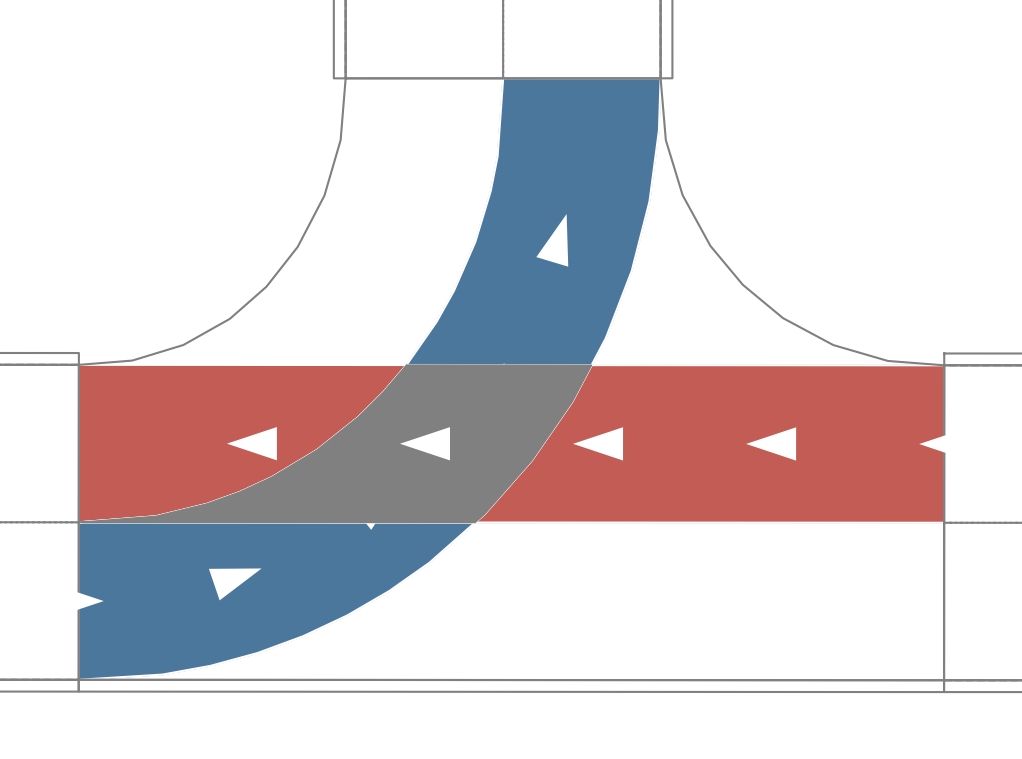} &
                \includegraphics[valign=c,width=\linewidth-0.5\tabcolsep]{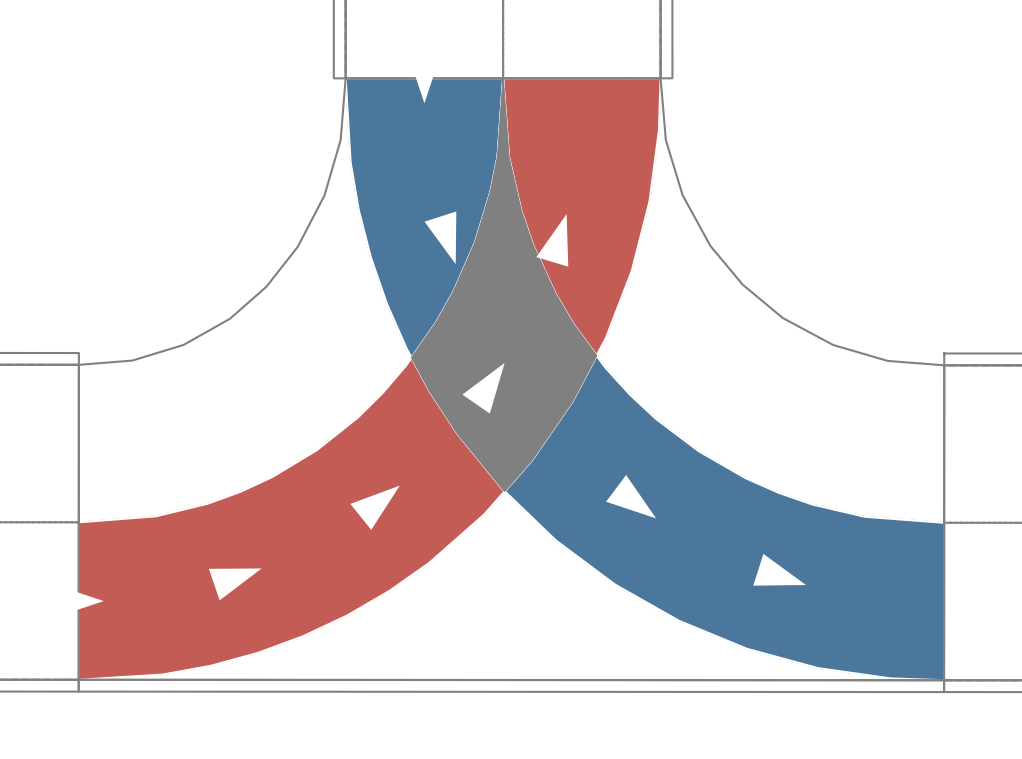} \\
                \hline
            \end{tabularx}
            \caption[]{Collision-inducing 2-actor scenarios \\ \phantom{placeholder} \\ \phantom{placeholder}}
            \label{fig:regions-2-actor}
        \end{subfigure}
        \begin{subfigure}[b]{0.23\textwidth}
            \centering
            \begin{tabularx}{\linewidth}{|X|}\hline
                \includegraphics[valign=c,width=\linewidth-0.5\tabcolsep]{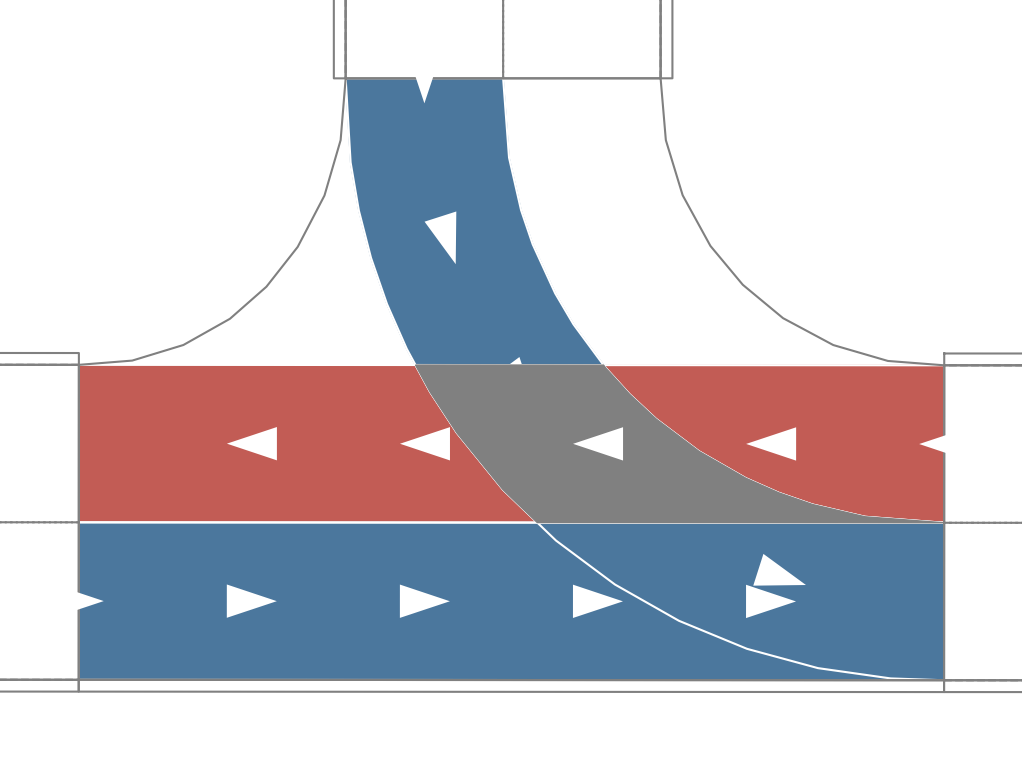} \\
                \includegraphics[valign=c,width=\linewidth-0.5\tabcolsep]{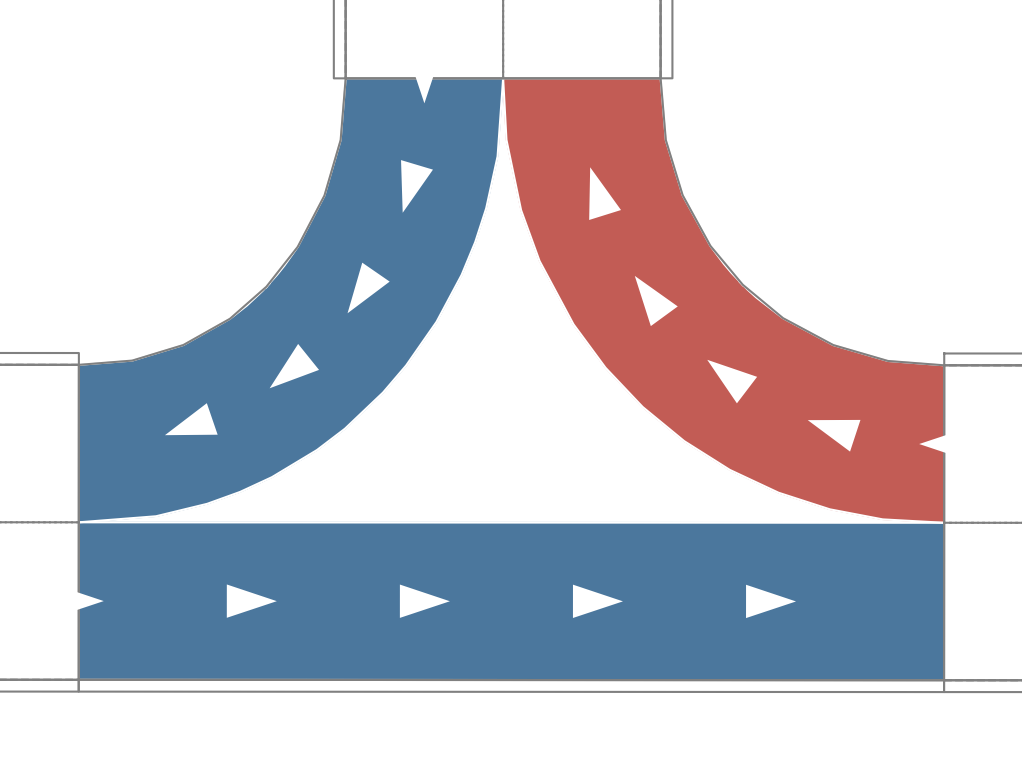} \\
                \hline
            \end{tabularx}
            \caption[]{3-actor scenarios not considered as dangerous}    
            \label{fig:regions-3-actor-invalid}
        \end{subfigure}
    \caption{Logical scenarios at a road junction. The designated path region of the ego and external actors (with arrows depicting their orientation) are depicted in red and in blue, respectively. Overlapping regions are depicted in gray.}
    \label{fig:regions}
\end{figure}

\begin{exline}
    \autoref{fig:regions} depicts logical scenarios at a 3-way, 1-lane road junction, which includes six possible maneuver instances (shown in \autoref{fig:regions-1-actor}).
    For 2-actor scenarios, there are 36 possible permutations of the six possible maneuvers, 24 of which satisfy the danger criterion proposed in this paper.
    Six of the 24 possible dangerous scenarios are shown in \autoref{fig:regions-2-actor}, where the designated path regions of the ego and external actors are depicted in red and blue, respectively. Their overlapping region is depicted in gray.
    
    Furthermore, we depict two dangerous 3-actor scenarios in \autoref{fig:regions-3-actor-valid}.
    Note that it is not necessary for a pair of external actors to have overlapping paths.
    We also depict two 3-actor scenarios in \autoref{fig:regions-3-actor-invalid} that are not considered to be dangerous (hence they cannot be generated by our approach), where at least one external actor has a designated path region that does not overlap with the designated path region of the ego actor.
    
    In such non-dangerous scenarios, there may be cases where some, but not all, external actors have a designated path region that overlaps with that of the ego actor.
    However, in this paper, we consider such scenarios as equivalent, in terms of potential danger, to lesser-actor scenarios where the non-overlapping external vehicle is excluded.
    For instance, the first 3-actor scenario in \autoref{fig:regions-3-actor-invalid} is not considered to be dangerous as it is subsumed by the first (dangerous) 2-actor scenario in \autoref{fig:regions-2-actor}.
\end{exline}

    

\newcommand{\numActors}[0]{$n_{ac}$}
\newcommand{\setManeuvers}[0]{$S_{\maninst{}}$}

\newcommand{\mapCurrentAssignments}[0]{$M_{cur}$}
\newcommand{\setScenarios}[0]{$S_{d}$}
\newcommand{\nextId}[0]{$id_{nxt}$}

\newcommand{\recursiveFunctionname}[0]{findLogScenarios}
\newcommand{\recursiveFunction}[0]{\textsc{\recursiveFunctionname}}

\newcommand{\maneuverOne}[1]{$man_{#1}$}
\newcommand{\id}[1]{$id_{#1}$}

\algrenewcommand\algorithmicrequire{\textbf{Input:}}
\algrenewcommand\algorithmicensure{\textbf{Output:}}


\begin{algorithm}
	\caption{Deriving Dangerous Logical Scenarios 
 }
	\begin{algorithmic}[1]
            \Require{\jut: junction under test}
            \Require{\numActors $> 1$: number of actors} 
            \Ensure{\setScenarios: set of dangerous logical scenarios}

            \State \textbf{global} \setManeuvers $\leftarrow$ \textsc{getAllManeuverInstances}(\jut)
            \State \Return \recursiveFunction($\varnothing$, $\varnothing$)
            \State

            \Procedure{\recursiveFunctionname}{\mapCurrentAssignments, \setScenarios}
                \If{size[\mapCurrentAssignments] $<$ \numActors }
                
                    \Comment{\textit{General case}: some actors are not assigned a maneuver}
                    
                    \State \nextId $\leftarrow$ size[\mapCurrentAssignments]
                    \For{\maneuverOne{} $in$ \setManeuvers}
                        \State \mapCurrentAssignments[\nextId] $\leftarrow$ \maneuverOne{}
                        \State \setScenarios{} $\leftarrow$ \recursiveFunction(\mapCurrentAssignments, \setScenarios)
                        \State \mapCurrentAssignments{} $\leftarrow$ \mapCurrentAssignments{} $\setminus \{$\nextId$\}$
                    \EndFor
                    \State \Return \setScenarios
                \Else
                
                    \Comment{\textit{Base case}: all actors are assigned a maneuver}
                    
                    \State \maneuverOne{ego} $\leftarrow$ \mapCurrentAssignments[0]
                    \For{\id{ext} in 1..\numActors}
                        \State \maneuverOne{ext} $\leftarrow$ \mapCurrentAssignments[\id{ext}]
                        \If{$\neg$ \textsc{areOverlapping}(\maneuverOne{ego}, \maneuverOne{ext})}

                            \State \Return \setScenarios
                            
                        \EndIf
                    \EndFor
                    \State \Return \setScenarios $\leftarrow$  \setScenarios $\bigcup \{$\mapCurrentAssignments$\}$
                \EndIf
            \EndProcedure
	\end{algorithmic} 
\label{algo:func-to-log}
\end{algorithm}

\textbf{Interface}: \autoref{algo:func-to-log} details our logical scenario generation approach.
It takes as input the number \numActors{} of required actors, which include one ego actor and \numActors$-1$ external actors, as well as the junction under test \jut.
The junction \jut{} contains a collection of possible \maninst{}s.
Our approach recursively derives the complete set \setScenarios{} of all dangerous logical scenarios over the collection of possible \maninst{}s at \jut.
In this context, a logical scenario is represented as a mapping between actors and their designated \maninst{}. 

\textbf{Initialization and function call}:
Our algorithm first derives the set \setManeuvers{} of all possibles \maninst{}s at the junction-under-test, according to its  underlying road map geometry.
It then calls the recursive \recursiveFunction{} function which takes as argument the current (possibly partial) mapping \mapCurrentAssignments{} between actors and their designated \maninst{}s.
This mapping is initially empty (i.e. an actor has not yet been assigned an \maninst{}).
It is built up during the general case and evaluated during the base case of the recursive algorithm.
\recursiveFunction{} takes as arguments the set \setScenarios{} of dangerous logical scenarios to be returned (which is initially empty).


In \textbf{the general case [L5..12]}, the algorithm builds up the \mapCurrentAssignments{} mapping between actors and \maninst{}s.
If, at a given stage, not all actors have been assigned an \maninst{} (size[\mapCurrentAssignments] $<$ \numActors), our approach first identifies the next actor \nextId{} to be assigned an \maninst{}, 
then iterates through all possible \maninst{}.
At each iteration, a new \maninst{} is assigned to the \nextId{} actor.
\recursiveFunction{} is then called recursively, with the newly expanded \mapCurrentAssignments{} given as input.
After completion of the recursive call, the \maninst{} assigned to the \nextId{} actor is retracted and a new \maninst{} is assigned to it.
Once all possible \maninst{}s have been iteratively assigned to the \nextId{} actor, the \textbf{for} loop terminates.
The general case thus ensures that all possible permutations for \maninst{} assignments are derived and evaluated in the base case.


\textbf{The base case [L14..22]} is reached when all actors have been assigned an \maninst{} (size[\mapCurrentAssignments] $==$ \numActors{}), i.e. when a logical scenario has been derived.
At this stage, our algorithm extracts the \maninst{} \maneuverOne{ego} assigned to the ego actor
and iterates over all \maninst{}s \maneuverOne{ext} assigned to external actors to evaluate whether they overlap with \maneuverOne{ego}.
If any \maneuverOne{ext} does not overlap with \maneuverOne{ego}, the logical scenario is not considered to be dangerous, which causes the base case to terminate and to return an unchanged \setManeuvers.
Otherwise, the current logical scenario is considered to be dangerous: it is appended to \setManeuvers, and the newly expanded \setManeuvers{} is returned.


\textbf{Extending path regions}:
Once each actor is assigned an \maninst{} (and a corresponding path region), we
extend the designated path regions with their (recursive) predecessors (i.e. the starting lane) and successors (i.e. the end lane), as shown in the bottom-left cell of \autoref{fig:scenarioAbsLevels}.
As such, the entire concrete path would be included in the extended path region designated to each actor at the logical-level.


\textbf{Completeness}:
We show that our proposed scenario generation approach satisfies the following completeness property:

\theorembox{1}{
For \numActors{} actors and a set \setManeuvers{} of possible maneuver instances at a junction \jut{} under test, \autoref{algo:func-to-log} derives the complete set of logical scenarios (maneuver instance assignments to actors) that satisfy the logical-level danger criterion that all external actors have overlapping path regions with the ego actor.}

By construction, the general case of \autoref{algo:func-to-log} iterates through all possible  permutations of \maninst{} assignments to actors, which defines the entire search space of the algorithm.
At this stage, the ego and external actors are treated identically.
Consequently, in the base case, every permutation is evaluated to determine whether or not it satisfies the danger criterion.
A permutation is included in the algorithm output if and only if it satisfies the danger criterion.




\section{Dangerous concrete scenarios}
\label{sec:Concrete}

\newcommand{\prevman}[0]{PM}

Once the set of all dangerous logical scenarios is found for a given \jut~and \numActors, we systematically refine each logical scenario into a dangerous concrete scenario,
where each actor is assigned a \textit{concrete path} (i.e. a sequence of exact points to follow).
As required by the functional-level specification, the generated concrete scenarios are collision-inducing.
During scenario execution, this requires the ego actor to perform \textit{preventive maneuvers} (\prevman{}s) to avoid collision.
In other words, if the ego actor follows its assigned path without considering external actors, it is expected to get into a collision.

\subsection{Deriving concrete paths}

\newcommand{\mapAssignments}[0]{$M_{log}$}
\newcommand{\mapSpeedProfiles}[0]{$M_{pro}$}
\newcommand{\initialPosition}[1]{$startpoint_{#1}$}
\newcommand{\mapFollowPoints}[0]{$M_{con}$}

\newcommand{\startlane}[1]{$startLane_{#1}$}
\newcommand{\ndlane}[1]{$endLane_{#1}$}
\newcommand{\maneuver}[1]{$man_{#1}$}
\newcommand{\concretepath}[1]{$path_{#1}$}
\newcommand{\concretepathpoint}[1]{$p_{#1}$}

\newcommand{\distToCollision}[1]{$distToColl_{#1}$}
\newcommand{\subpathToCollision}[1]{$spth_{#1}$}
\newcommand{\timeToCollision}[1]{$t_{#1}$}
\newcommand{\timeonly}[1]{$t_{#1}$}
\newcommand{\remainingTime}[1]{$remainingTime_{#1}$}
\newcommand{\listAssignments}[1]{$L_{#1}$}

\newcommand{\pathregion}[1]{$r_{#1}$}
\newcommand{\overlapRegion}[1]{\pathregion{ovl}}

\newcommand{\pointalonglogicalpath}[2]{\textsc{#1PointAlongPathRegion}(\pathregion{#2})}
\newcommand{\getpointin}[0]{\textsc{getPointIn}}

\begin{algorithm}
	\caption{Deriving Concrete Path Assignments
 }
	\begin{algorithmic}[1]
            \Require{\mapAssignments: mapping from actors to designated logical-level \maninst{}s}
            \Require{\mapSpeedProfiles: mapping from actors to assigned speed/accel. profiles}
            \Ensure{\mapFollowPoints: mapping from actors to assigned concrete paths}
            

            \Comment{Find ego-actor path}

            \State \pathregion{ego} $\leftarrow$ \mapAssignments[0].pathRegion

            \State \concretepathpoint{ego} $\leftarrow$ \getpointin(\pathregion{ego}.predecessor)
            \State \concretepath{ego} $\leftarrow$ \textsc{List}(\concretepathpoint{ego})
            \While {$\neg$ \concretepathpoint{ego} $\in$ \pathregion{ego}.successor}
                \State \concretepathpoint{ego} $\leftarrow$ \concretepathpoint{ego}.\pointalonglogicalpath{next}{ego}
                \State \concretepath{ego}.append(\concretepathpoint{ego})
            \EndWhile
            \State \mapFollowPoints[0] $\leftarrow$ \concretepath{ego}

            \Comment{Find external-actor paths}

            \State \listAssignments{\maninst{}} $\leftarrow$ \textsc{sort}( [($k$, $v$.pathRegion) $\mid$ $\exists k \neq 0, v$ : ($k$, $v$) $\in$ \mapAssignments{}] ) 


            \For{ (\id{ext},\pathregion{ext}) in \listAssignments{\maninst{}}}
                
                \Comment{Get ego actor time to collision}
                
                \State \overlapRegion{ext} $\leftarrow$ \textsc{getOverlapRegion}(\pathregion{ego}, \pathregion{ext})

                \State \subpathToCollision{ego} $\leftarrow$ \textsc{subpathToRegion}(\concretepath{ego}, \overlapRegion{ext} )
                \State \timeToCollision{ego} $\leftarrow$ \textsc{pathTime}(\subpathToCollision{ego}, \mapSpeedProfiles[0])

                \Comment{Get external actor starting point}

                \State \timeToCollision{ext} $\leftarrow$ \timeToCollision{ego} $+$ \textsc{getTimePenalty}(\listAssignments{\maninst{}})

                \State \concretepathpoint{ext} $\leftarrow$ \getpointin(\overlapRegion{ext})

                \State \concretepath{ext} $\leftarrow$ List(\concretepathpoint{ext})
                \While {\timeToCollision{ext} $>$ 0}
                    \State \concretepathpoint{pre} $\leftarrow$ \concretepathpoint{ext}.\pointalonglogicalpath{prev}{ext}
                    \State \subpathToCollision{ext} $\leftarrow$ \textsc{pathBetween}(\concretepathpoint{pre}, \concretepathpoint{ext})
                    \State \timeonly{sub} $\leftarrow$ \textsc{pathTime}(\subpathToCollision{ext}, \mapSpeedProfiles[\id{ext}])

                    \State \concretepath{ext}.prepend(\concretepathpoint{pre})
                    \State \timeToCollision{ext} $\leftarrow$ \timeToCollision{ext} $-$ \timeonly{sub}
                    \State \concretepathpoint{ext} $\leftarrow$ \concretepathpoint{pre}
                \EndWhile
                \State \textsc{extendFowardAlongPathRegion}(\concretepath{ext}, \pathregion{ext})
                \State \mapFollowPoints[\id{ext}] $\leftarrow$ \concretepath{ext}
            \EndFor


	\end{algorithmic} 
 \label{algo:log-to-conc}
\end{algorithm}

\textbf{Overview}: 
\autoref{algo:log-to-conc} details our proposed concrete scenario generation approach.
It takes as input a dangerous logical scenario \mapAssignments{} (a mapping between actors and their designated \maninst{}s), as derived by \autoref{algo:func-to-log}.
Additionally, our approach takes as input a mapping \mapSpeedProfiles{} that assigns to each actor a speed/acceleration profile which determines their temporal behavior during scenario execution.
\mapSpeedProfiles{} is integrated as an external component that may vary between case studies in accordance to specific testing objectives.
For instance, such a profile may assign target speeds to different road types (e.g. 40km/h on residential roads, 100 km/h on highways), with acceleration values during transitions. 
Such profiles may also be a function of time (as opposed to road types).

Our approach first derives a concrete path designated for the ego actor along the direction of its designated path region (derived from its \maninst{}).
Next, for each external actor, it finds the overlapping region between its designated path region and that of the ego actor.
It then builds up the external actor path by extending backwards from the overlapping region.
The actor speed/acceleration profiles are taken into consideration to ensure that generated paths are collision-inducing.
Our approach outputs a mapping \mapFollowPoints{} between actors and their designated concrete path.

\textbf{The ego actor path [L1..8]}:
As an initial step, we determine the path region \pathregion{ego} designated to the ego actor from its designated \maninst{} and select a starting point (i.e. the initial value of \concretepathpoint{ego}) for the ego actor.
This is some point in the region preceding \pathregion{ego} (i.e. its \textit{predecessor}) which is determined by a call to some (customizable) \getpointin() function.
We build up the concrete path \concretepath{ego} from this starting point by successively appending the \textit{next} path point (determined by following the direction of \pathregion{ego} for a certain distance.) until we reach the target region (the \textit{successor} of \pathregion{ego}).

\textbf{The external actor paths [L9..27]}:
According to \autoref{algo:func-to-log}, all \maninst{}s assigned to an external actor overlap with the \maninst{} assigned to the ego actor. 
As a preparatory step, we derive the list \listAssignments{\maninst{}} of external actor path regions sorted in order of overlap along the ego path (i.e. if the ego actor is expected to collide with actor A before colliding with actor B, actor A is placed earlier in the sorted list).


For each external actor with id \id{ext} and path region \pathregion{ext}, we first find the overlap region \overlapRegion{ext} between the designated path regions for the ego and \id{ext} actors.
For the ego actor, we extract its (concrete) sub-path \subpathToCollision{ego} between its starting point and \overlapRegion{ext}.
Note that the exact end of the sub-path may be any point in or around \overlapRegion{ext}, decided upon according to some (custom) criterion.
We then determine the time \timeToCollision{ego} required by the ego actor to reach \overlapRegion{ext} as a function of its associated speed/acceleration profile \mapSpeedProfiles[0].

\newcommand{\coll}[1]{$c_{#1}$}
For each external actor, we generate a path such that it reaches \overlapRegion{ext} at the same time as the ego actor (to cause a collision).
However, in cases where another potential collision \coll{A} is planned to occur before the collision with actor \id{ext},
we expect the ego actor to perform some \prevman{} in response to \coll{A}, which would delay its arrival to \overlapRegion{ext}.
As such, the expected time to collision \timeToCollision{ext} for the \id{ext} external actor is the sum of \timeToCollision{ego} and an approximate time penalty to account for a \prevman{} performed in response to \coll{A}.
Such a time penalty varies according to case-study-specific parameters such as actor speeds and junction geometry.

Once \timeToCollision{ext} is determined, we select a target point \concretepathpoint{ext} for the external actor in \overlapRegion{ext} (according to some \getpointin() function) and we incrementally move backwards to build up its concrete path.
At each iteration of the while loop, we derive the previous point \concretepathpoint{pre} along the assigned path region and we determine the travel time \timeonly{sub} between the two points.
Once the time budget is exhausted (i.e. \timeToCollision{ext} $\geq$ 0), the start point of the external actor is reached (i.e. \concretepath{ext} contains the path between the start point and \overlapRegion{ext}).
To complete \concretepath{ext}, we extend it past \overlapRegion{ext} along the path region for some distance.
The exact distance is irrelevant for testing purposes since the potential collision would have already occurred.

An important property of our algorithm is as follows:

\theorembox{2}{
For any pair of intersecting paths of the ego and external actors at a concrete-level scenario, there exist overlapping path regions for the respective actors at the logical level. 
}

Thus, each collision-inducing logical scenario serves as an equivalence class for a (potentially infinite) set of concrete scenarios that may result in a collision during simulation. Thanks to the completeness of our logical scenario generation algorithm (see \autoref{algo:func-to-log}), our approach provides coverage for each equivalence class. 

\subsection{Simulating concrete scenarios}

\textbf{Static analysis}: 
It is possible that for a given maneuver assignment and speed/acceleration profiles, the derived scenario cannot be adequately simulated (e.g. in cases where the initial positions of two actors overlap). 
Furthermore, there may be cases where external actors would collide with each other before the potential ego-actor collision.
Considering that our derived paths correspond to a specific speed/acceleration profile, such cases may be detected and handled \textit{prior to scenario execution} through static analysis.
All other cases (that pass such static checks) are eligible for simulation.

\textbf{Dynamic analysis}:
Despite the relevance of static analysis, the exact behavior of actors in simulation cannot be predicted a priori, particularly for autonomous agents controlled by a neural network.
For instance, our proposed approach assigns an approximate time penalty (in L14 of \autoref{algo:log-to-conc}) to account for possible \prevman{}s performed by the ego actor.
The precision of such a time penalty can only be evaluated at runtime, through simulation. 

Considering this uncertainty, the use of such approximations in deriving concrete scenarios may result in unintended interactions between the ego and external actors.
For instance, dynamic analysis may show that a recorded collision involving the ego actor could not have been avoided through ego-actor \prevman{}s.
In the context of autonomous sensor-based agents, we consider a collision with an external object to be \textit{avoidable} only if the autonomous agent perceives (without necessarily identifying) the colliding object through its sensors for a long enough time prior to collision.

\newcommand{\prevfigwidth}[0]{0.47\textwidth}
\begin{figure}
    \centering     
        \begin{subfigure}[b]{0.47\columnwidth}
            \centering
            \includegraphics[width=\prevfigwidth]{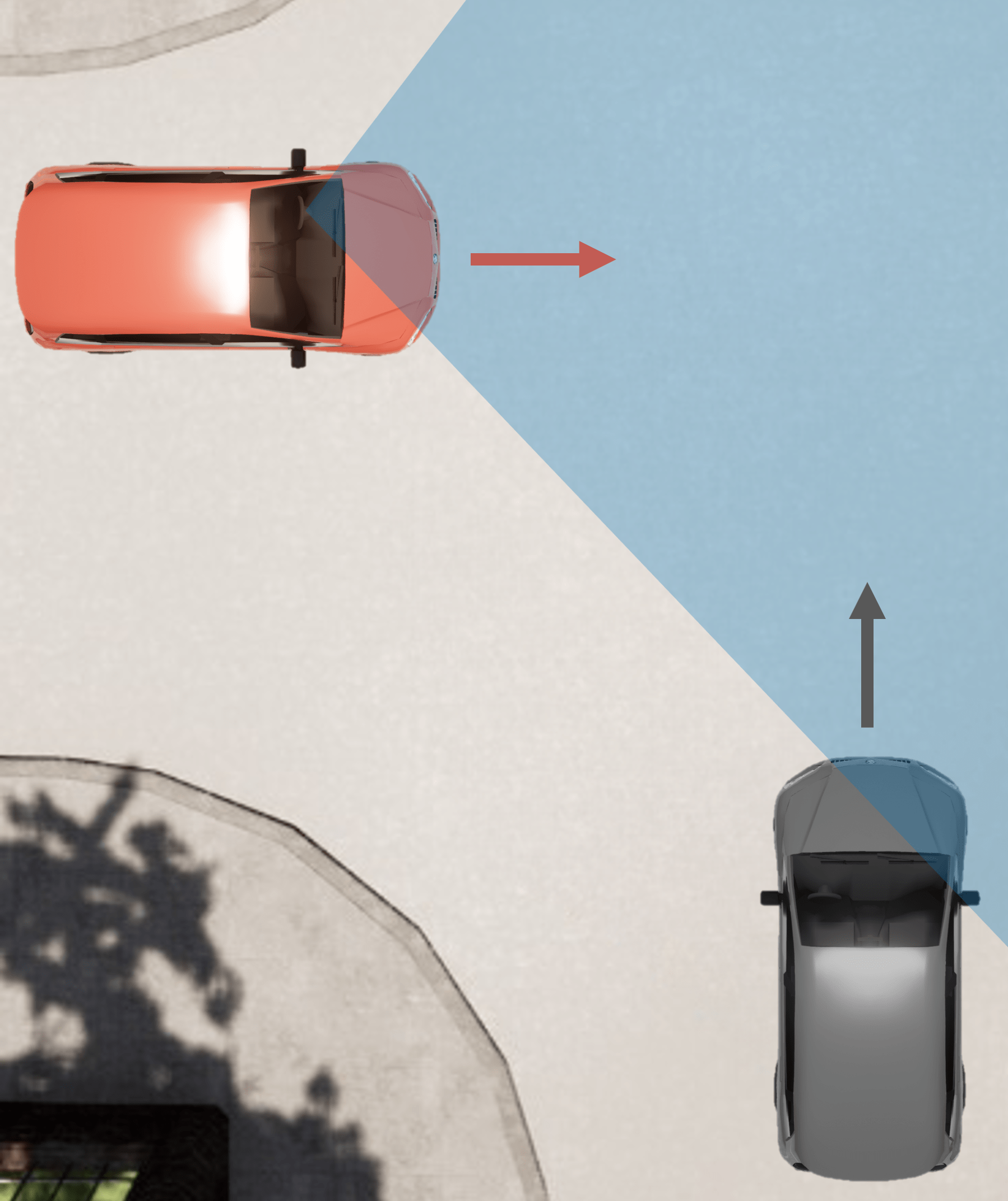}
            \hfill
            \includegraphics[width=\prevfigwidth]{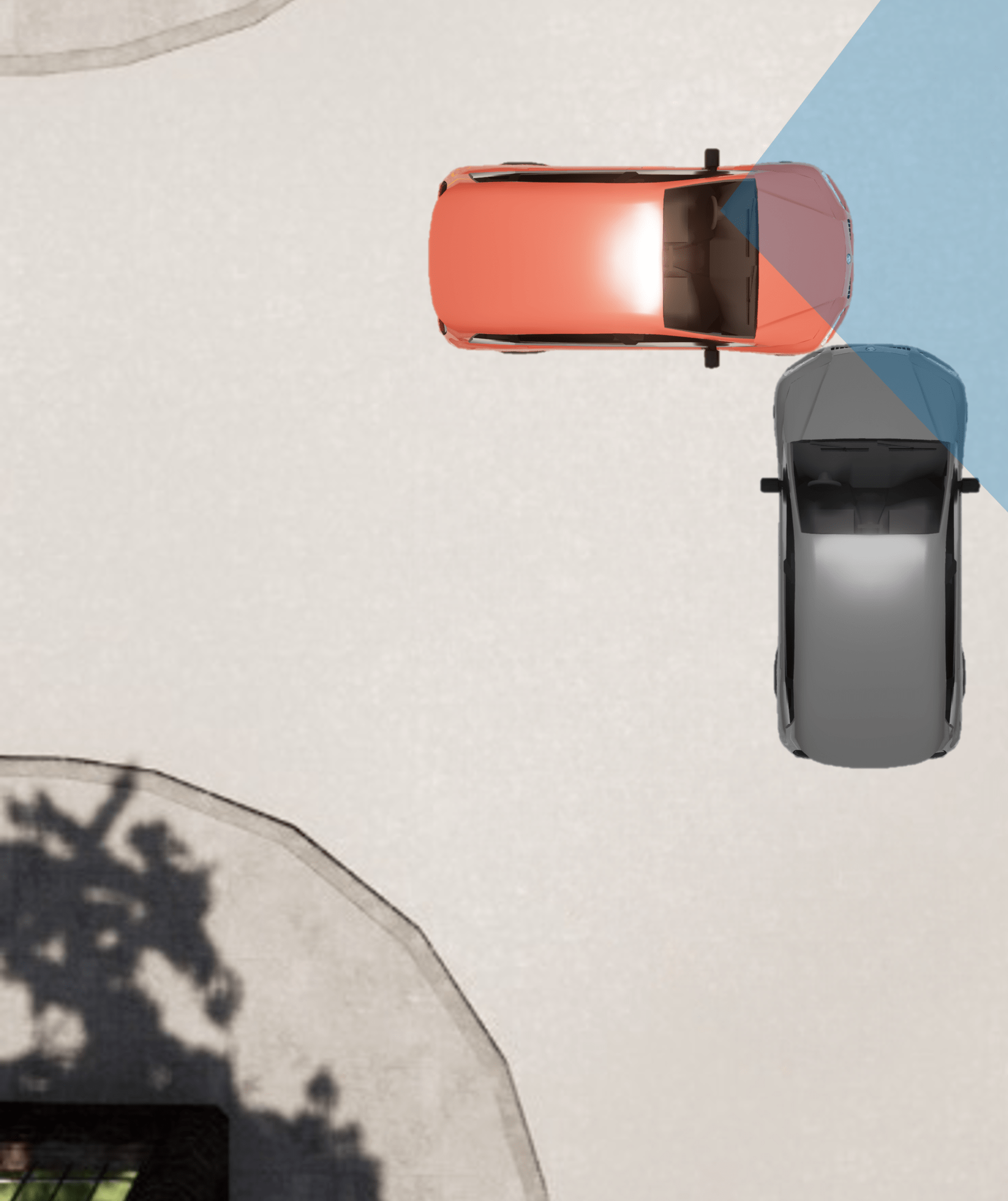}
            \caption[]{Avoidable collision - longer detected pre-collision period }
            \label{fig:avoidable-coll}
        \end{subfigure}
        \hfill
        \begin{subfigure}[b]{0.47\columnwidth}
            \centering
            \includegraphics[width=\prevfigwidth]{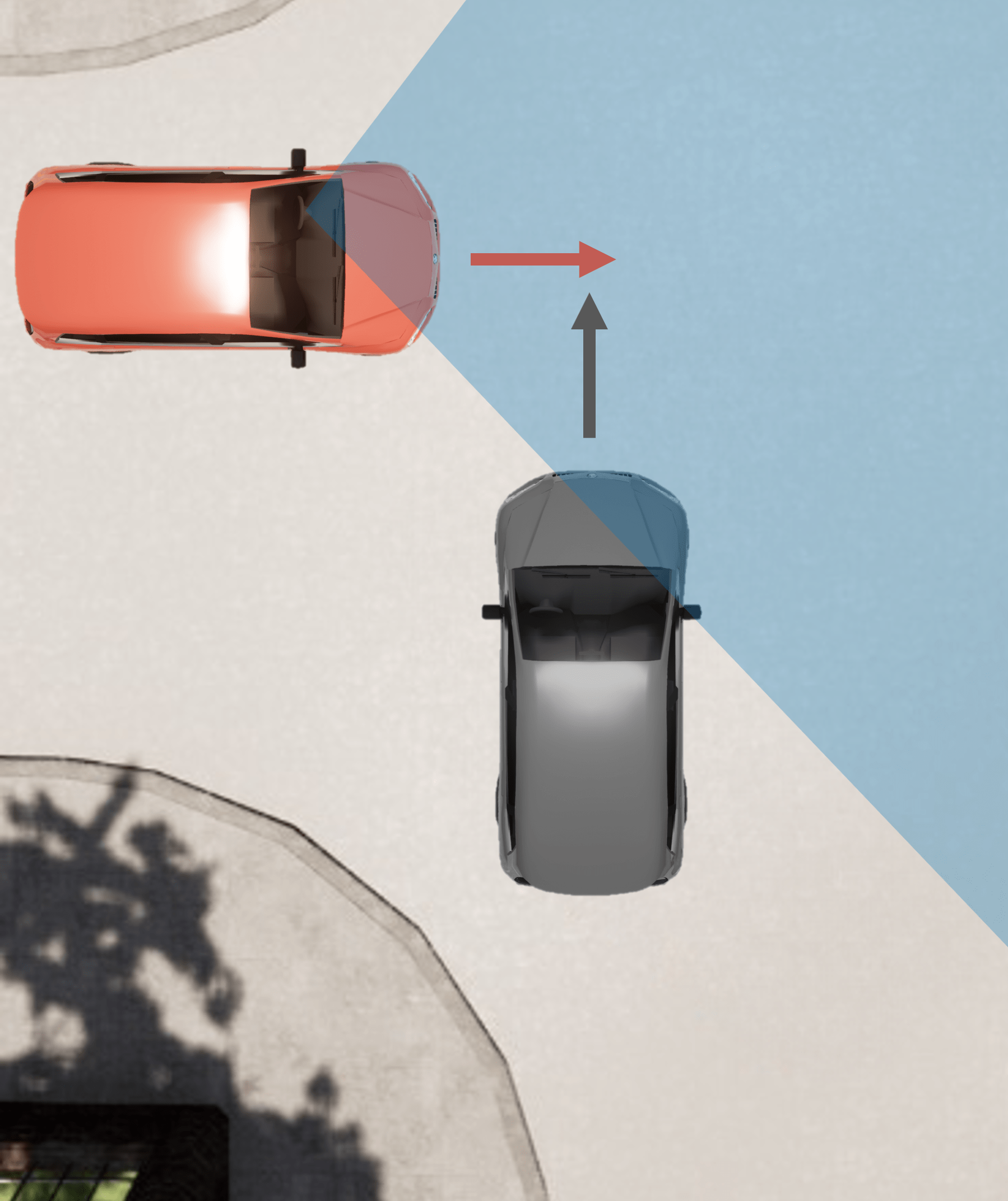}
            \hfill
            \includegraphics[width=\prevfigwidth]{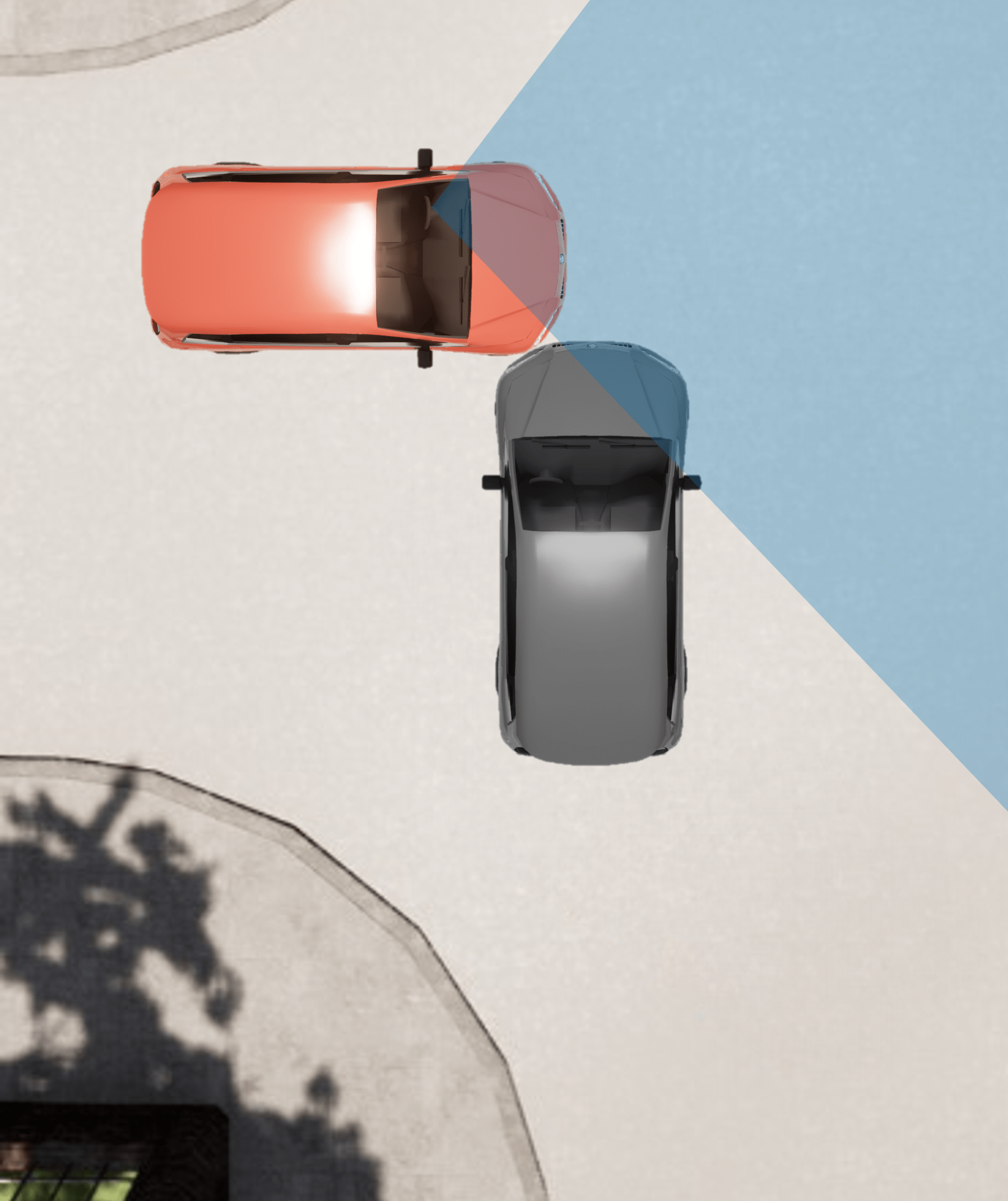}
            \caption[]{Unavoidable collision - shorter detected pre-collision period }
            \label{fig:unavoidable-coll}
        \end{subfigure}
    \caption{Scenes depicting the moment of  detection (left) and of collision (right) for avoidable and unavoidable collisions.}
    \label{fig:both-collisions}
\end{figure}




\begin{exline}
    \autoref{fig:both-collisions} shows two colliding simulation runs.
    The ego actor and the external actor are shown in red and grey, respectively.
    The ego actor's visibility sector, detected through a front facing camera attached to the AV, is depicted in blue.
    Actor velocities are identical for all actors and are represented as arrows.
    
    For both collisions, we depict the simulation state (1) at the initial moment where the ego actor detects the external actor, and (2) at the moment of their collision.
    In \autoref{fig:avoidable-coll}, the time between the initial detection and the collision is long enough for the ego actor to potentially perform a \prevman{}.
    This makes the collision depicted in \autoref{fig:avoidable-coll} avoidable.
    However, this is not the case for the unavoidable collision depicted in \autoref{fig:unavoidable-coll}: the detection period is too short and the collision cannot be avoided.

\end{exline}

\textbf{Functional-level compliance}:
During scenario generation, a derived concrete scenario is a specification that is fed into the simulator framework.
By construction, \textit{a derived concrete scenario specification} (i.e. the concrete path assignment) is consistent wrt. to the original functional scenario given as input.
At simulation time, i.e. during scenario execution, the ego vehicle may deviate from its assigned concrete path as a response to some external safety threat (see \autoref{sec:pre-scene} for details).
As such, while scenario execution may deviate from the original functional scenario, (completeness and compliance) properties of the generation approach are not altered.

\section{Evaluation}
\label{sec:evaluation}


Our experimental evaluation addresses the following research questions.

\begin{itemize}
  \item \rquestion{1}: How often does our approach yield dangerous situations in simulation without considering unavoidable collisions?
  
  \item \rquestion{2}: How does the level of danger vary with the increasing number of external actors?
  \item \rquestion{3}: How does the level of danger depend on functional-level maneuvers?
  \item \rquestion{4}: How do road geometry and scenery affect the level of danger for the same functional maneuver?

\end{itemize}

\rquestion{1} justifies the relevance of our scenario generation approach by evaluating whether our approach can generate collision-inducing scenarios with preventable maneuvers (i.e. scenarios that are neither trivial, nor unavoidable).
\rquestion{2}, \rquestion{3} and \rquestion{4} contribute to the broader research objective of reliably (i.e. with empirical evidence) measuring and comparing the danger level provided by specific parameters or concepts that define scenarios at various abstraction levels.
Specifically, \rquestion{2} and \rquestion{3} evaluate the influence of functional-level parameters (number of actors and maneuvers, respectively), while \rquestion{4} compares assigned (logical) path regions.

\newcommand{\juncRur}[0]{\textsc{Rural}}
\newcommand{\juncUrb}[0]{\textsc{Urban}}

\begin{figure}
    \centering
        \begin{subfigure}[b]{0.47\columnwidth}
            \centering
            \includegraphics[width=\linewidth]{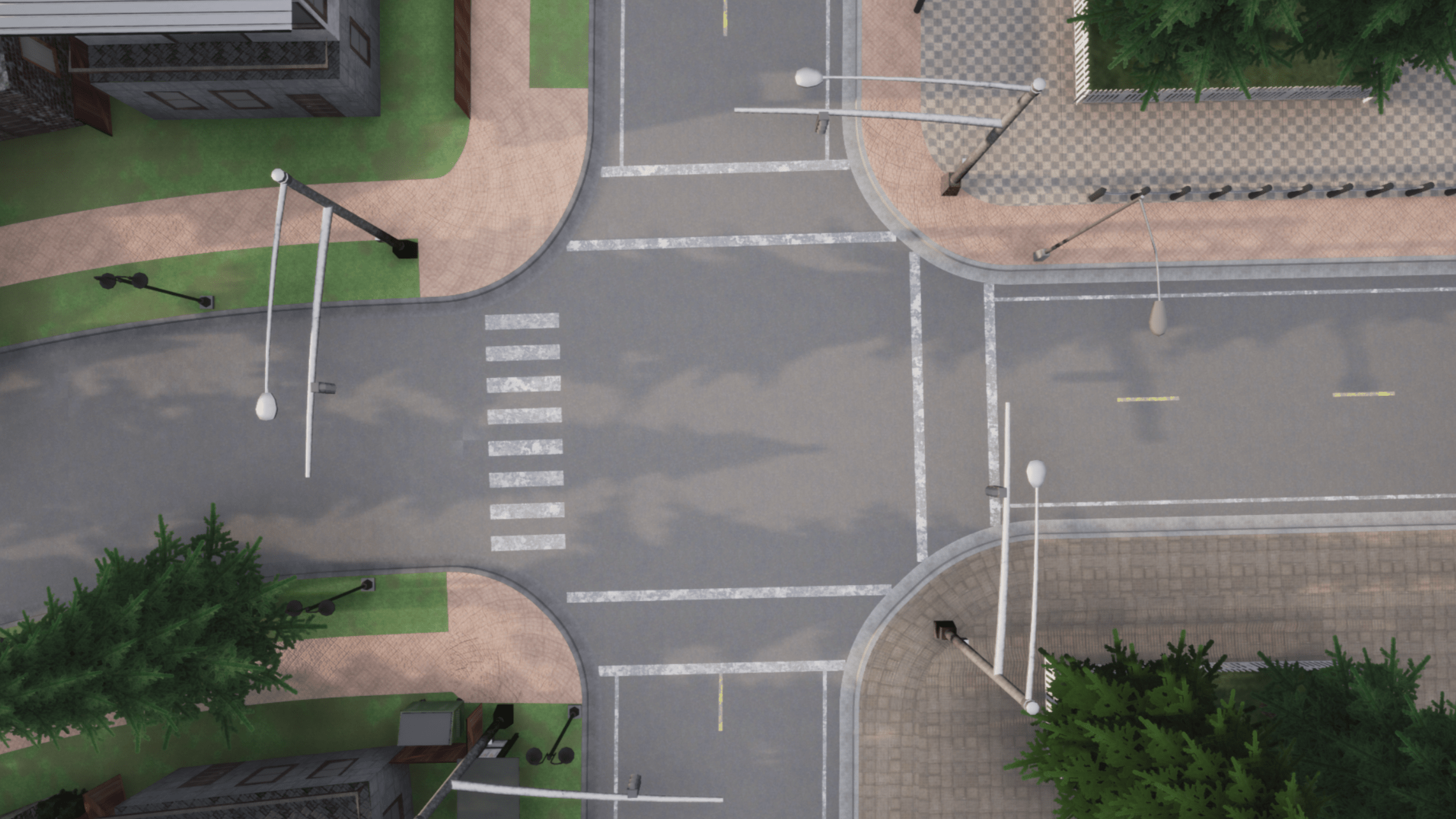}
            \caption[]{\juncRur{} - in simulation}
            \label{fig:rural-screenshot}
        \end{subfigure}
        \hfill
        \begin{subfigure}[b]{0.47\columnwidth}
            \centering
            \includegraphics[width=\linewidth]{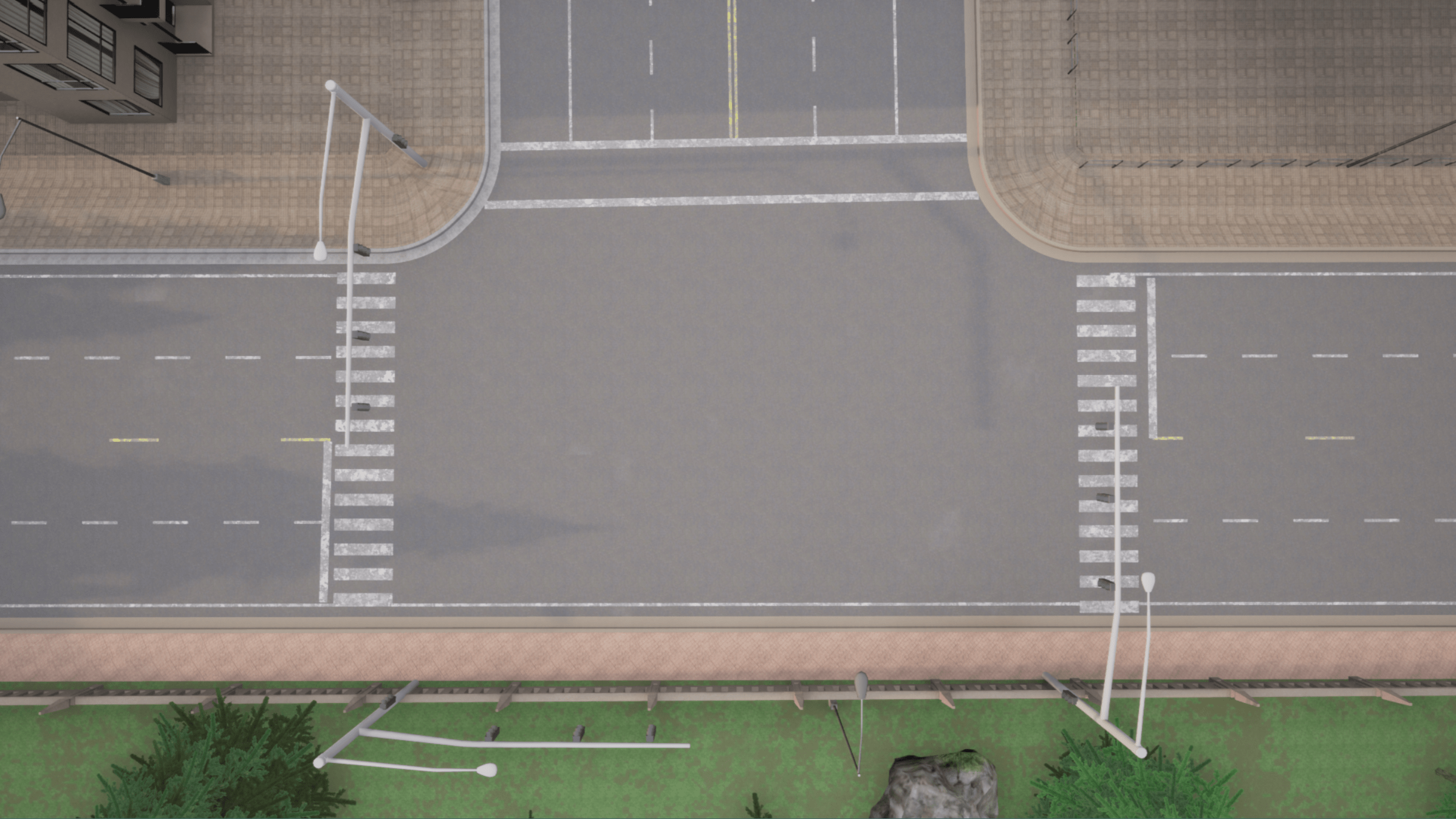}
            \caption[]{\juncUrb{} - in simulation}
            \label{fig:urban-screenshot}
        \end{subfigure}
    \caption{The road junctions under test}
    \label{fig:junctions}
\end{figure}

\subsection{Road junctions}
To answer these research questions, we generate, simulate, and evaluate test traffic scenarios over 2 road junctions with diverse characteristics included in the CARLA simulator framework \citep{Dosovitskiy2017CarlaSimulator}.
An illustration of each road junction is presented in \autoref{fig:junctions}.

\juncRur{} is a 4-way road junction with one incoming and one outgoing lane at each connecting road segment.
It is included in the \textit{Town 4} road map of CARLA, which depicts a small town with a backdrop of snow-capped mountains and conifers.
Despite its geometric symmetry, the junction is surrounded by diverse scenery elements: the SW and NW corners contain small houses with prominent conifers, the NE corner contains a tall apartment building, while the SE corner contains a medium-sized shopping mall.

The junction contains 4 instances (one per connecting road segment) of each functional maneuver (\qa{turnLeft}, \qa{turnRight} and \qa{goStraight}), for a total of 12 logical path regions.

\juncUrb{} is a 3-way road junction with two incoming and two outgoing lanes at each connecting road segment.
It is included in the \textit{Town 5} road map of CARLA, which depicts an urban environment with a backdrop of conifer-covered hills.
The scenery surrounding the road junction includes skyscrapers at both corners of the junction and a conifer forest at its south.

The junction contains a total of 8 logical path regions (associated to 2 \qa{turnLeft}, 2 \qa{turnRight}, and 4 \qa{goStraight} maneuvers).

\textbf{AV agent:}
As part of our experiments, we perform safety analysis of the Transfuser \citep{Prakash2021CVPR,Chitta2023PAMI} sensor-based AV controller, which has been evaluated as a case study in existing research \citep{Haq2023Reinforcement}.
Transfuser is mounted with a LiDAR (which considers points within a 32m $\times$ 32m square in front of the vehicle) and three cameras (which consider points within a front-facing 132$^{\circ}$ FOV).
We use this information to address the avoidability of recorded collisions.

\subsection{General measurement setup}


In our experiments, we generate (regular) test scenarios where actors follow road lanes.
We exclude corner cases where actors follow illegal paths (e.g. off-road paths).
As such, logical-level path regions directly correspond to \textit{real lanes} in the road map.
At the concrete level, exact paths designated to actors follow the center-line of the corresponding path region, which represents the \textit{most typical concretization} (in terms of expected lawful driver behavior).
To increase interactions between actors at the junction-under-test, actors are programmed to disregard road signals during simulation.


\begin{table}[htb]
    \centering
    \caption{Number of derived (A) logical scenarios, (B) scenarios w/o symmetries, (C) scenarios w/o initial overlaps, per road junction, per scenario size. Notation: (A)$\rightarrow$(B)$\rightarrow$(C)}
    \begin{tabular}{l|ccc}
        \hline
         Junction  & 2 actors & 3 actors & 4 actors \\
        \hline
         \juncRur & 92$\rightarrow$92$\rightarrow$56 & 748$\rightarrow$420$\rightarrow$124 & 6332$\rightarrow$1460$\rightarrow$160 \\
         \juncUrb & 26$\rightarrow$26$\rightarrow$14 & 102$\rightarrow$64$\rightarrow$13 & 134$\rightarrow$32$\rightarrow$6 \\
        \hline
    \end{tabular}
    \label{tab:num_gen_scen}
\end{table}

\textbf{Scenario generation:}
We first derive the set of \emph{all possible collision-inducing logical scenarios} with an increasing number of actors, for up to 4 actors (including the ego actor).
We set target scenario sizes aligned with the scalability offered by many recent publications \citep{Riccio2020,Abdessalem2018TestingVisionBased,Abdessalem2016TestingADAS,Haq2023Reinforcement,Zhong2021Fuzzing,Abdessalem2018TestingFeatureInteractionsBriandNsgaForConcreteScenes,calo2020GeneratingAvoidableCollision,Wu2021,Klischat2019} that handle similar scenario generation challenges. 

The number of generated dangerous scenarios derived per road junction and scenario size is indicated in \autoref{tab:num_gen_scen}.
At this stage, we reduce the number of generated scenarios by removing symmetric cases where the same combination of \maninst{}s is assigned to the external actors.
For instance, consider two three-actor scenarios $S_A$ and $S_B$. In both cases, the ego actor is assigned $\maninst{}_1$.
In $S_A$, the external actors $ext_1$ and $ext_2$ are respectively assigned $\maninst{}_2$ and $\maninst{}_3$.
In $S_B$, they are respectively assigned $\maninst{}_3$ and $\maninst{}_2$.
In such a case, $S_A$ and $S_B$ are considered to be symmetric.


We then derive concrete scenarios by assigning the default speed profile of Transfuser (i.e. 3 m/s inside a junction and 4 m/s outside a junction) to all actors.
Considering the identical speed profiles assigned to all actors, symmetric scenarios are equivalent for simulation purposes.
During scenario generation, we include a time penalty for prior potential collisions as the sum of the time required for the prior colliding external actor to traverse the overlapping region and a 1-second penalty to account for ego actor acceleration 


Among the generated concrete scenarios, some cannot be adequately simulated as they require the initial positions of actors to overlap.
For instance, since actor speeds are identical in our experiments, such a case arises when two actors are assigned the same \maninst{}.
However, this is not a conceptual limitation of our approach as it can handle arbitrary speed and acceleration profiles given as input.
The final number of derived concrete scenarios per junction, per scenario size, is indicated in \autoref{tab:num_gen_scen}.


\textbf{Simulation:}
We execute each generated scenario 10 times within the CARLA simulator framework \citep{Dosovitskiy2017CarlaSimulator}.
We use the default vehicle model (Tesla Model 3) and color (dark blue) for external actors.
We use default weather and lighting conditions throughout.
We conduct our evaluation on a \textit{g4dn.2xlarge} AWS EC2 instance with eight virtual cores, NVIDIA T4 GPU (16GB), 32 GB memory, running Ubuntu 20.04.

In our experiments, both the generation (a few seconds) and the execution (up to 30 seconds per run) of test scenarios terminated in reasonable time.  
While more complex road junctions and a higher number of actors increase the number of test scenarios, their execution (in simulation) can be scaled out using our cloud-based infrastructure, hence the end-to-end analysis time is controllable.  



\textbf{Analysis:}
We consider a simulation run to be dangerous if it results in a collision, a near-miss or if the ego actor performs a \prevman{} to avoid a collision.
For cases where a collision occurs, we evaluate whether the collision was avoidable.
We also evaluate the statistical significance of our results through the \textit{Fisher exact test} \citep{Fisher1992FishersExactTest} (for p-value) and \textit{odds ratio} \citep{Haddock1998OddsRatio} (for effect size), as suggested in existing guidelines \citep{Arcuri2011} for comparing dichotomous outcomes (i.e. success or near-miss/collision, \prevman{}s performed or not).

\textit{Collisions} are detected at runtime by a dedicated sensor attached to the ego actor.
A \textit{near-miss} occurs if no collision is detected,
but an external actor gets closer to the ego actor than a predefined distance  (1 unit/meter) at some stage.

Considering that our approach focuses on deriving logical scenarios from functional scenarios, we provide measurement analysis at abstract (functional and logical) levels.
Conversely, existing scenario generation approaches often solely focus on concretizing logical scenarios given as input, and thus provide concrete-level measurement analysis.
Due to these discrepancies, we cannot use existing approaches as baseline for our evaluation. 

\textbf{Avoidable collisions:}
To determine if a collision is avoidable, we measure whether the external actor colliding with the ego actor is detected by the ego-actor sensor(s) for a certain period of time during scenario execution.
At a given point in time during scenario execution, an external actor is considered to be detected by (a) the camera or (b) the LiDAR sensor if at least one part of the external actor is within (a) its visibility sector, or (b) its detection area.

In our experiments, a collision is considered to be avoidable if the colliding external actor is detected by both the camera and the LiDAR sensor during at least 90\% of the 60 frames directly prior to the collision.
This corresponds to 2.7s of the 3 seconds before collision, which, at the speed of 3 m/s, is a reasonable amount of time for the AV to react to any dangerous situations.
In our analysis, we exclude simulation runs that contain unavoidable collisions.

\textbf{Preventive maneuvers:}
To determine whether a \prevman{} was performed by the ego actor in response to external actors during simulation, we first determine the exact trajectory that the ego actor would have traversed if there were no external actors.
To do so, we remove the external actors and run the scenario (10 times).
Among the 10 monitored paths, we select as reference a medoid path that has the smallest maximal distance to any other path.
By manual inspection, we also justify our assumption of \autoref{sec: overview} stating that the ego actor does not deviate outside of its designated path region when no dangerous situation is encountered.

Once the reference path for the ego actor is determined, we reinclude the external actors, run the scenario in simulation and compare the monitored ego actor path to the reference path.
For each frame in the simulation, we associate it to a point on the reference path that is the closest to the ego position at that frame.
If we detect a point in the reference path that associates to more frames than its predecessor point (with a certain minimal threshold, in our case, 5 frames), this indicates that the ego actor has performed a preventive slow-down maneuver near that point. 

\textbf{Evaluation artifacts:} All artifacts derived as part of our experimental evaluation are included in a dedicated publication page\footnote{\url{https://doi.org/10.5281/zenodo.13286656}}. Artifacts include generated logical and concrete scenarios, as well as simulation traces and analysis figures.


\subsection{RQ1: Danger and preventive maneuvers}
\label{sec: RQ1}

\renewcommand\tabularxcolumn[1]{m{#1}}
\begin{figure}[htp]

    \centering
    \includegraphics[width=0.75\linewidth]{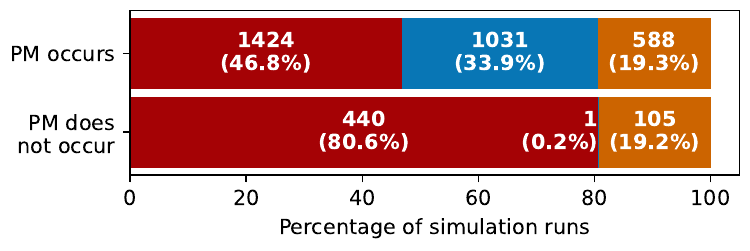}
    \includegraphics[width=.6\linewidth]{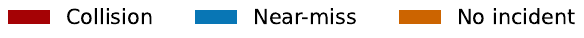}
    \caption{Scenario execution outcomes wrt. \prevman{} occurrence. Scenarios containing unavoidable collisions are disregarded. The label on each bar depicts the number (and percentage) of simulation runs that showcase the corresponding outcome.}
    \label{fig:both-prev-outcomes}
\end{figure}

\noindent
\textit{In this experiment, we evaluate (1) whether the ego actor performs \prevman{}s (i.e. whether generated scenarios are not trivial) and (2) whether these \prevman{}s are sufficient for collision avoidance (i.e. whether danger induced by our generated scenario is avoidable).}

\textbf{Dangerous simulations:}
\autoref{fig:both-prev-outcomes} shows the aggregate distributions (over all scenario sizes) of simulation outcomes (collision, near-miss, no incident) according to the presence of \prevman{}s.
From 3730 simulation runs, we include results for 3589 runs: 141 runs (3.7\%) containing unavoidable collisions are disregarded.
From the 3589 runs, 3043 (84.8\%) included \prevman{}s.
From the remaining 546 runs where no \prevman{} occurs, 441 runs (12.3\%) either resulted in a collision or a near-miss.

Our results show that 97.1\% (3484/3589) of simulation runs include a dangerous situation (i.e. a \prevman{}, a collision or a near-miss).
Furthermore, analysis shows that among the 373 generated scenarios, only 6 (1.6\%) did not result in any dangerous simulation runs.
Note that the ratio of dangerous simulation runs is not only dependent on the scenario itself, but also on the various thresholds used during analysis (due to the non-deterministic nature of the AV controller).
For instance, using a smaller near-miss threshold could result in some of the observed near-misses to be considered as no-incident cases instead.

\textbf{Preventive maneuvers:}
\autoref{fig:both-prev-outcomes} further shows that regardless of \prevman{}s, the ratio of no-incident simulations remains constant, at slightly over 19\%.
In cases including \prevman{}s, collisions, and near-misses occur for 46.8\% and 33.9\% of simulation runs, for a total of 80.7\%.
When no \prevman{}s occur, 80.6\% of simulation runs result in a collision.
From this, we conclude that the \prevman{}s performed by the ego actor do promote its safety by avoiding collisions and causing near-misses instead for 33.9\% of simulation runs.

\ranswer{1}{97.1\% of simulation runs over all scenario sizes result in a dangerous situation. Furthermore, 84.8\% of dangerous runs include \prevman{}s performed by the ego actor.
These \prevman{}s are sufficient to avoid collisions for 33.9\% of runs with PMs and cause near-misses instead.}


\newcommand{\tablefigwidth}[0]{0.32\textwidth}
\newcommand{\tablefigsmallwidth}[0]{0.2\textwidth}
\begin{figure}[htp]

  \begin{tabular}{
    |>{\centering\arraybackslash}m{\dimexpr.4\linewidth-2\tabcolsep-1.3333\arrayrulewidth}
    |>{\centering\arraybackslash}m{\dimexpr.3\linewidth-2\tabcolsep-1.3333\arrayrulewidth}
    |>{\centering\arraybackslash}m{\dimexpr.3\linewidth-2\tabcolsep-1.3333\arrayrulewidth}|
    }
    
    \rquestion{2} & \multicolumn{2}{c|}{\rquestion{3}}\\
    & \juncUrb{} & \juncRur{}\\
    
    \includegraphics[width=\linewidth]{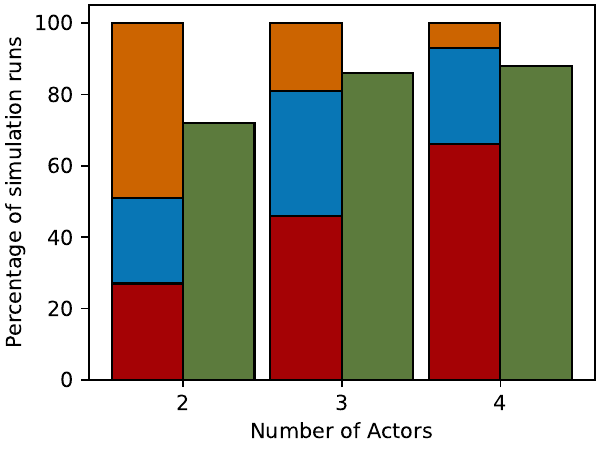}
    & \includegraphics[width=\linewidth]{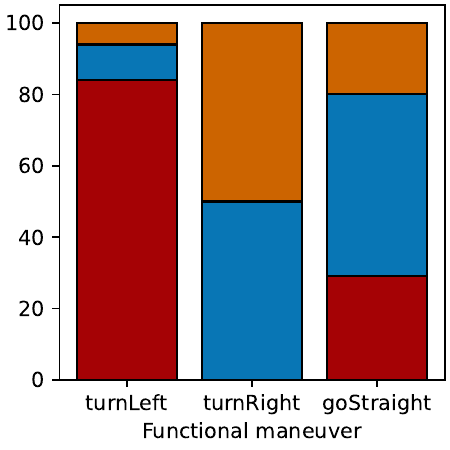}
    & \includegraphics[width=\linewidth]{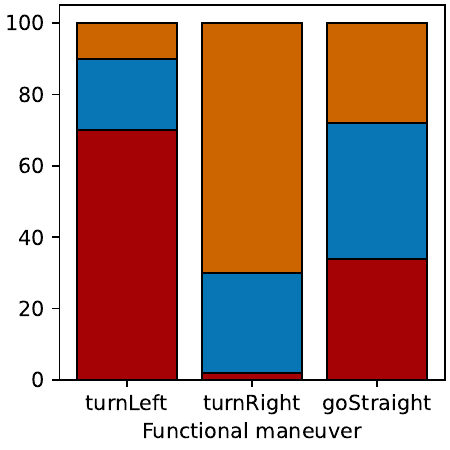}
    \\
   
  \end{tabular}

\centering
  \begin{tabular}{
    |>{\centering\arraybackslash}m{\dimexpr.3\linewidth-2\tabcolsep-1.3333\arrayrulewidth}
    |>{\centering\arraybackslash}m{\dimexpr.3\linewidth-2\tabcolsep-1.3333\arrayrulewidth}|
    }
  
    \multicolumn{2}{|c|}{\rquestion{4}}\\
    \juncUrb{} & \juncRur{}\\
    
    \includegraphics[width=\linewidth]{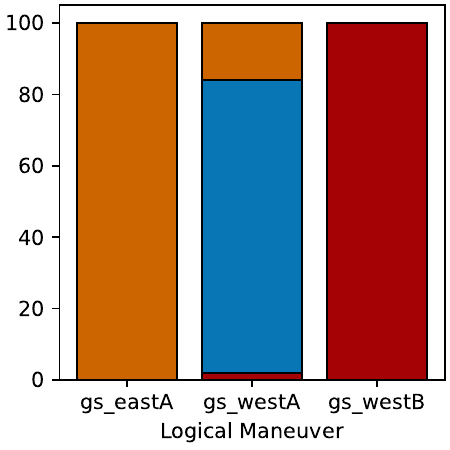}
    & \includegraphics[width=\linewidth]{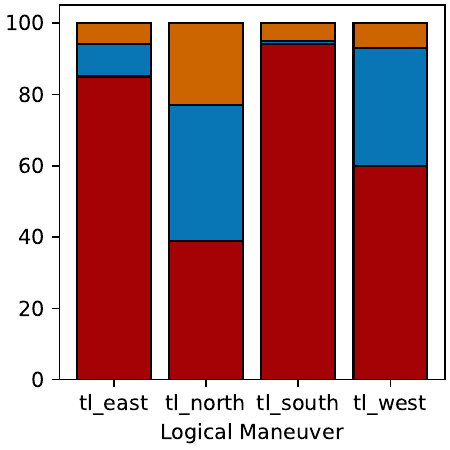}\\
  
  \end{tabular}

  \includegraphics[width=\linewidth]{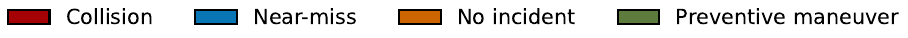}
  
  \captionsetup{justification=centering}
  \caption{Measurement data for scenario execution outcomes wrt. the increasing number of actors, (\rquestion{2}), functional-level maneuver assignments to ego (\rquestion{3}) and logical level path maneuvers assigned to ego (\rquestion{4})}
\label{fig:eval-rq1}
\end{figure}

\subsection{RQ2: Increasing number of external actors}
\label{sec: RQ2}

\textit{This experiment aims to determine whether the presence of additional external actors (which is a functional parameter) increases the level of danger provided by a scenario.}
---
The top-left subfigure of \autoref{fig:eval-rq1} shows the simulation outcomes of runs (aggregated for both road junctions) alongside the presence of \prevman{}s with an increasing number of scenario actors.
Our result shows that the ratio of simulations with unsafe outcomes (collisions and near-misses) increases with scenario size.
In particular, the percentage of collisions increases from 27.9\% for one external actor to 66.7\% for 3 external actors.
Furthermore, the percentage of \prevman{}s also increases from 72.9\% to 88.4\%.
This is an intuitive result: a simulation run is more likely to have an unsafe outcome (and, in response, include \prevman{}s), when it encounters more collision-inducing situations.


\ranswer{2}{Increasing the number of external actors results in a higher ratio of (1) unsafe simulation outcomes and (2) preventive maneuvers. In particular, the percentage of colliding simulations increases from 27.9\% for one external actor to 66.7\% for three external actors.}


\subsection{RQ3: Functional-level maneuver analysis}
\label{sec: new-RQ3}

\textit{This experiment aims to detect whether different ego actor (functional-level) maneuver assignments provide different levels of danger.}
---
The center and right subfigures at the top of \autoref{fig:eval-rq1} shows the simulation outcomes according to the functional-level maneuver assigned to the ego actor for \juncUrb{} and \juncRur{}, respectively, aggregated for all scenario sizes.
The ego actor is assigned the \qa{turnLeft}, \qa{turnRight} and \qa{goStraight} functional maneuvers, respectively, for 1595, 114 and 1550 simulation runs for the \juncRur{} junction and for 210, 10, 110 simulation runs for the \juncUrb{} junction.

In both cases, we detect a similar (intuitive) trend regarding simulation outcomes:
scenarios involving a left turn are most dangerous (94.8\% and 90.3\% of simulation runs for \juncUrb{} and \juncRur{}, respectively, involve a collision or a near-miss).
Conversely, right-turn scenarios are least dangerous (no incident occurs for 50.0\% and 69.3\% of simulation runs for \juncUrb{} and \juncRur{}, respectively, and collision rates are below 2.7\% for both cases).

For the \juncRur{} junction, both conclusions are statistically significant ($p < 0.05$).
Effect size varies between 1.341 (small) and 26.683 (large) for the various measurements.
For the \juncUrb{} junction, our conclusions are similar, but they are not supported by statistical significance measurements. 

\ranswer{3}{Functional-level analysis shows that scenarios where the ego actor needs to perform a left turn are most dangerous (over 90.0\% of runs result in an unsafe outcome). Conversely, right-turn scenarios are the least dangerous (with a collision rate of under 2.7\%).}


\subsection{RQ4: Road geometry and scenery analysis}
\label{sec: new-RQ4}


\textit{This experiment aims to detect if, for the same assigned functional-level maneuver, the level of danger varies according to the road geometry and scenery (wrt. the assigned logical path region).}
---
We consider all scenarios where the ego actor is assigned the same functional maneuver, categorize them by logical-level \maninst{} assignment, and compare the simulation outcomes.

For the \juncUrb{} junction, we compare the \qa{goStraight} \maninst{} assignments since they associate to the most number of path regions.
We particularly consider the geometric properties of these path regions wrt. other interacting lanes.
We evaluate three path regions: \qa{gs\_eastA} (10 simulations, with a single incoming left-turn path region), \qa{gs\_westA} (70 simulations, with one incoming right turn and two crossing left turns), and \qa{gs\_westB} (30 simulations, with two crossing left-turn regions).
Results are shown in the bottom-left subfigure of \autoref{fig:eval-rq1}. 
On one hand, \qa{gs\_eastA} assignments succeed for all simulations: it results in less unsafe situations with statistical significance ($p<0.05$) and large effect size than both other \maninst{} assignments.
On the other hand, \qa{gs\_westB} assignments collide for all simulations: they result in more collisions with statistical significance ($p<0.05$) and large effect size.
\qa{gs\_westA} assignments predominantly result in near-misses.

We attribute these differences in outcomes to the specific approach angles and path regions assigned to external actors.
The ego actor better avoids collision when external actors follow an incoming lane (as opposed to a crossing lane).
We attribute these results to the training of the underlying AV-under-test: Transfuser has been trained in regular traffic, where (1) cut-in maneuvers are common (particularly outside of road junctions), and (2) external actors follow road signals (cases where an external actor crosses quasi-perpendicularly in front of the ego actor are rare).

Considering the geometric symmetry of the \juncRur{} junction, we compare the \qa{turnleft} \maninst{} assignments since they involve the most simulation runs.
Specifically, each path region assignment results in 400 simulation runs (minus 5 total cases for unavoidable collisions).
Results in the bottom-right subfigure of \autoref{fig:eval-rq1} show that east- and south-bound left turns cause collisions more often (85.7\% and 94.8\%, respectively) than the north- and west-bound equivalents (39.4\% and 60.8\%, respectively).
This conclusion is statistically significant ($p < 0.05$), with medium effect size, between 1.411 and 2.405.

The geometric symmetry of the road junction entails the structural symmetry of logical scenarios.
As a result, we attribute the variance in scenario outcomes to the differences in scenery elements surrounding the road junction, being the only differences between simulation runs from the ego actor's perspective.


\ranswer{4}{For the \juncUrb{} junction, the differences in road geometry result in different dominating simulation outcomes for the three examined \qa{goStraight} \maninst{}s. For the symmetric \juncRur{} junction, we notice a difference of at least 24.9\% in the collision frequency between the safest and least safe pair of left-turn \maninst{}s.}

\subsection{Limitations and threats to validity}


\textbf{Construct validity.}
Our approach guides scenario generation using a reasonable multi-level representation of danger based on potential collisions.
At the logical level, we mitigate construct validity by taking as input the set of possible maneuver instances (and corresponding path regions) at a road junction.
At the concrete level, we make various reasonable approximations, i.e. related to time penalties for prior potential collisions and exact in-simulation behavior of AVs, during scenario generation.
To mitigate the effect of these approximations, we address their precision during evaluation and subject them to strict validation conditions (i.e. relating to the avoidability of collisions and the presence of \prevman{}s).
During analysis, we also rely on preliminary measurements to determine reasonable thresholds relating to near-misses and to vehicle detection time for avoidable collisions.

\textbf{Internal validity.}
We mitigate threats to internal validity by deriving a complete set of logical scenarios.
Since this paper does not focus on logical-to-concrete scenario refinement, we propose a deterministic approach that (consistently) refines each derived logical scenario into a \textit{single} concrete scenario.
To increase the relevance of concrete scenarios, we derive the \textit{most typical case} (in terms of expected lawful driver behavior) by integrating center-line following.
Although our proposed concrete scenario derivation approach may miss certain variants of a collision, these variants are abstractly incorporated within path regions, thus do not lead to \textit{new potential collisions} at the logical level.
Nevertheless, we do reduce the influence of exact concretization parameters by performing functional- and logical-level aggregations during analysis.
As future work, we may extend our approach to derive a set of similar, but not identical, concrete scenarios from a given logical scenario.


\textbf{External validity.}
We mitigate threats to external validity by performing measurements over two road junctions with different characteristics.
This allows us to evaluate 373 logical scenarios, which is orders of magnitude greater than the evaluation proposed in recent related publications \citep{Riccio2020,Abdessalem2018TestingVisionBased,Abdessalem2016TestingADAS,Haq2023Reinforcement,Zhong2021Fuzzing,Abdessalem2018TestingFeatureInteractionsBriandNsgaForConcreteScenes,calo2020GeneratingAvoidableCollision,Wu2021,Klischat2019}.
In line with observed practices in recent publications \citep{Riccio2020, Haq2022, luLearningConfigurationsOperating2023}, we evaluate (in simulation) the behavior of a real AV controller used in existing scenario generation research.
We perform thorough evaluation (a total of 3730 simulation runs) to adequately, and consistently, address all derived logical scenarios.
We accompany our measurement with statistical significance analysis in accordance with best practices to mitigate the differences in the number of generated scenarios for each junction.




\section{Related Work}
\label{sec:relWork}

\newcommand{\li}[1]{\textit{(#1)}}

\noindent\textbf{Abstract scenario specification.}
Existing scenario specification languages often build upon a conceptual basis for traffic scenarios proposed by \citet{Ulbrich2015Defining}, \citet{Menzel2018ScenariosForDevelopment} and \citet{steimle2021}.
These approaches describe the various components and abstraction levels required for scenario specification.
At a conceptual level, scenarios are often defined through multi-layer representations \citep{Schuldt2018,Bagschik2018,Scholtes2021,Urbieta2021} which provide a hierarchy for traffic scenarios.


Those approaches define functional scenarios with a custom specification, that is tailored to represent a specific type of scenario (focusing on a single maneuver).
As such, they lack in expressiveness and extensibility.
An exception is \scenic~\citep{Fremont2019ScenicLanguage}, which handles arbitrary scenario specifications as input, even though it has some expressiveness limitations. Ontologies \citep{Geyer2014,Klueck2018,Bagschik2018,Urbieta2021} can provide a formal basis for an extensible functional scenario specification.

Other specification approaches use \textit{temporal concepts} as building blocks for scenarios.
Such temporal concepts include (1) reasoning over vehicle paths \citep{Queiroz2019GeoScenarioAn}, (2) sequence of vehicle behaviors \citep{Schutt2020} or (3) conditional state transitions between scenes \citep{Hempen2017}. The expressivity and extensibility of these approaches are limited, 
but initial concretization results are often provided. More expressive scenario specification languages have also been proposed using model-based implementations~\citep{Bach2016ModelBasedScenario} and a state-based representation adapted for search-based scenario generation~\citep{Hempen2017}.
However, case studies that use these scenario description languages are 
limited to the derivation of simplistic or fixed scenarios.


Scenario description languages may also focus on the concrete representation of scenarios, through temporal behavior defined by paths attributed of dynamic components \citep{Queiroz2019GeoScenarioAn} and through a grid representation adapted to machine learning applications \citep{Ries2019}.
In both cases, the proposed specification languages may only be used as an extension of the output language proposed in this paper.

\noindent\textbf{Scenario concretization approaches.}
\textit{Search-based approaches} 
are most commonly used for scenario generation.
Many-objective search can be used to test feature interactions in AVs \citep{Abdessalem2018TestingFeatureInteractionsBriandNsgaForConcreteScenes}, to perform efficient \textit{online} testing \citep{Haq2022} and to address the branch coverage of test suite generation approaches \citep{Panichella2015}.
Additionally, \citet{Abdessalem2016TestingADAS,Abdessalem2018TestingVisionBased} rely on multi-objective search  and learnable evolutionary algorithm to guide scenario generation towards \textit{critical scenarios}.
Critical scenarios have also been derived using a weighted search-based approach \citep{calo2020GeneratingAvoidableCollision} and genetic algorithm \citep{Wu2021}.
Similarly, \textsc{DeepJanus} \citep{Riccio2020} combines evolutionary and novelty search to derive test inputs at the behavioral frontier of AVs, while \citet{Babikian2021dReal} use a hybrid, graph and numeric solver-based approach to concretize a limited-visibility pedestrian crossing scenario.
Note that different search-based approaches may represent domain-specific constraints differently in the underlying algorithm \citep{Fan2017} (e.g. objectives vs. hard constraints). 

\textit{Sampling-based approaches} have also been used for scenario generation. 
Certain approaches \citep{Majumdar2021ParacsmJournal, Rocklage2018AutomatedScenarioGeneration} sample over a parametric (discrete) representation of arbitrary functional input scenarios, but they often avoid numeric (continuous) constraints (i.e. the logical scenario).
\citet{OKelly2018ScalableEndToEnd} use Monte Carlo sampling (over a continuous domain) to simulate scenarios with rare events, thus reasoning directly at the logical scenario level.
However, all these sampling-based approaches are limited to a fixed map location. 
The \scenic~framework \citep{Fremont2019ScenicLanguage} provides sampling-based concretization, but it improves on other sampling-based approaches by handling any road map as an input parameter.


\textit{Path-planning approaches} \citep{Althoff2018, Klischat2019} address scenario generation directly at the level of numeric constraints.
As such, they cannot handle abstract constraints as input.
These approaches often use formalizations of safety requirements as guiding metrics for scenario generation.
While, in principle, path-planning approaches are adaptable to road maps, no such experimental results are available.

\section{Conclusion}
\label{sec:conclusion}

In this paper, we propose an approach to generate dangerous traffic scenarios at a road junction, where danger is defined by the existence of potential collisions between actors. Our approach uses an abstract danger heuristic (i.e. overlapping path region assignments) to derive the complete set of avoidable, collision-inducing logical scenarios for functional level specifications given as input.
Logical scenarios are then refined into collision-inducing concrete scenarios amenable to simulation, where actors are tasked to follow exact paths along a speed and acceleration profile given as input.

We conduct \textbf{extensive experiments} to evaluate the behavior of a state-of-the-art learning-based AV controller on scenarios generated over two realistic road junctions with an increasing 
number of potentially colliding actors. 
Results show that the AV-under-test encounters increasing percentages of unsafe situations in simulation, which vary according to functional- and logical-level scenario properties.

As \textbf{future work}, we plan to extend our approach beyond one-maneuver scenarios (to include maneuver sequences) and to add support for conceptually complex danger definitions (e.g. traffic rule violations).
We also plan to use the insights provided by our results to determine more fine-tuned danger heuristics for guiding functional and logical scenario suite generation.

\section*{Acknowledgements}
We would like to thank Bal\'{a}zs Pint\'{e}r and Krist\'{o}f Marussy for their help in running the measurements.
This paper was partially supported by the Wallenberg AI, Autonomous Systems and Software Program (WASP), Sweden, by an Amazon Research Award and by the Department of Navy award (N629092412063) issued by the Office of Naval Research.

The authors have no competing interests to declare that are relevant to the content of this article.

% ...

\bibliography{bib/bib-mendeley}


\begin{thebibliography}{48}
\ifx \bisbn   \undefined \def \bisbn  #1{ISBN #1}\fi
\ifx \binits  \undefined \def \binits#1{#1}\fi
\ifx \bauthor  \undefined \def \bauthor#1{#1}\fi
\ifx \batitle  \undefined \def \batitle#1{#1}\fi
\ifx \bjtitle  \undefined \def \bjtitle#1{#1}\fi
\ifx \bvolume  \undefined \def \bvolume#1{\textbf{#1}}\fi
\ifx \byear  \undefined \def \byear#1{#1}\fi
\ifx \bissue  \undefined \def \bissue#1{#1}\fi
\ifx \bfpage  \undefined \def \bfpage#1{#1}\fi
\ifx \blpage  \undefined \def \blpage #1{#1}\fi
\ifx \burl  \undefined \def \burl#1{\textsf{#1}}\fi
\ifx \doiurl  \undefined \def \doiurl#1{\url{https://doi.org/#1}}\fi
\ifx \betal  \undefined \def \betal{\textit{et al.}}\fi
\ifx \binstitute  \undefined \def \binstitute#1{#1}\fi
\ifx \binstitutionaled  \undefined \def \binstitutionaled#1{#1}\fi
\ifx \bctitle  \undefined \def \bctitle#1{#1}\fi
\ifx \beditor  \undefined \def \beditor#1{#1}\fi
\ifx \bpublisher  \undefined \def \bpublisher#1{#1}\fi
\ifx \bbtitle  \undefined \def \bbtitle#1{#1}\fi
\ifx \bedition  \undefined \def \bedition#1{#1}\fi
\ifx \bseriesno  \undefined \def \bseriesno#1{#1}\fi
\ifx \blocation  \undefined \def \blocation#1{#1}\fi
\ifx \bsertitle  \undefined \def \bsertitle#1{#1}\fi
\ifx \bsnm \undefined \def \bsnm#1{#1}\fi
\ifx \bsuffix \undefined \def \bsuffix#1{#1}\fi
\ifx \bparticle \undefined \def \bparticle#1{#1}\fi
\ifx \barticle \undefined \def \barticle#1{#1}\fi
\bibcommenthead
\ifx \bconfdate \undefined \def \bconfdate #1{#1}\fi
\ifx \botherref \undefined \def \botherref #1{#1}\fi
\ifx \url \undefined \def \url#1{\textsf{#1}}\fi
\ifx \bchapter \undefined \def \bchapter#1{#1}\fi
\ifx \bbook \undefined \def \bbook#1{#1}\fi
\ifx \bcomment \undefined \def \bcomment#1{#1}\fi
\ifx \oauthor \undefined \def \oauthor#1{#1}\fi
\ifx \citeauthoryear \undefined \def \citeauthoryear#1{#1}\fi
\ifx \endbibitem  \undefined \def \endbibitem {}\fi
\ifx \bconflocation  \undefined \def \bconflocation#1{#1}\fi
\ifx \arxivurl  \undefined \def \arxivurl#1{\textsf{#1}}\fi
\csname PreBibitemsHook\endcsname

\bibitem[\protect\citeauthoryear{ISO}{2018}]{ISO26262FunctionalSatefy}
\begin{botherref}
\oauthor{\bsnm{ISO}}:
{ISO} 26262-1, road vehicles — functional safety.
International Organization for Standardization
(2018)
\end{botherref}
\endbibitem

\bibitem[\protect\citeauthoryear{ISO}{2019}]{ISO21448SOTIFRoadVehicles}
\begin{botherref}
\oauthor{\bsnm{ISO}}:
{ISO/PAS} 21448, road vehicles - safety of the intended functionality.
International Organization for Standardization
(2019)
\end{botherref}
\endbibitem

\bibitem[\protect\citeauthoryear{Czarnecki}{2018}]{Czarnecki2018Taxonomy}
\begin{botherref}
\oauthor{\bsnm{Czarnecki}, \binits{K.}}:
On-road safety of automated driving system - taxonomy and safety analysis
  methods.
Technical report,
U. of Waterloo
(2018)
\end{botherref}
\endbibitem

\bibitem[\protect\citeauthoryear{Majzik
  et~al.}{2019}]{Majzik2019TowardsSystemLevel}
\begin{bchapter}
\bauthor{\bsnm{Majzik}, \binits{I.}},
\bauthor{\bsnm{Semeráth}, \binits{O.}},
\bauthor{\bsnm{Hajdu}, \binits{C.}},
\bauthor{\bsnm{Marussy}, \binits{K.}},
\bauthor{\bsnm{Szatmári}, \binits{Z.}},
\bauthor{\bsnm{Micskei}, \binits{Z.}},
\bauthor{\bsnm{Vörös}, \binits{A.}},
\bauthor{\bsnm{Babikian}, \binits{A.A.}},
\bauthor{\bsnm{Varró}, \binits{D.}}:
\bctitle{Towards system-level testing with coverage guarantees for autonomous
  vehicles}.
In: \bbtitle{ACM/IEEE 22nd Int'l Conf. on Model Driven Engineering Languages
  and Systems, MODELS 2019},
pp. \bfpage{89}--\blpage{94}
(\byear{2019})
\end{bchapter}
\endbibitem

\bibitem[\protect\citeauthoryear{Kalra and Paddock}{2016}]{Kalra2016}
\begin{barticle}
\bauthor{\bsnm{Kalra}, \binits{N.}},
\bauthor{\bsnm{Paddock}, \binits{S.M.}}:
\batitle{Driving to safety: How many miles of driving would it take to
  demonstrate autonomous vehicle reliability?}
\bjtitle{Transportation Research Part A: Policy and Practice}
\bvolume{94},
\bfpage{182}--\blpage{193}
(\byear{2016})
\end{barticle}
\endbibitem

\bibitem[\protect\citeauthoryear{Koopman and Wagner}{2016}]{Koopman2016}
\begin{barticle}
\bauthor{\bsnm{Koopman}, \binits{P.}},
\bauthor{\bsnm{Wagner}, \binits{M.}}:
\batitle{Challenges in autonomous vehicle testing and validation}.
\bjtitle{SAE International Journal of Transportation Safety}
\bvolume{4}(\bissue{1}),
\bfpage{15}--\blpage{24}
(\byear{2016})
\end{barticle}
\endbibitem

\bibitem[\protect\citeauthoryear{Helle et~al.}{2016}]{Helle2016}
\begin{barticle}
\bauthor{\bsnm{Helle}, \binits{P.}},
\bauthor{\bsnm{Schamai}, \binits{W.}},
\bauthor{\bsnm{Strobel}, \binits{C.}}:
\batitle{Testing of autonomous systems - challenges and current
  state-of-the-art}.
\bjtitle{INCOSE International Symposium}
\bvolume{26}(\bissue{1}),
\bfpage{571}--\blpage{584}
(\byear{2016})
\end{barticle}
\endbibitem

\bibitem[\protect\citeauthoryear{Abdessalem
  et~al.}{2018}]{Abdessalem2018TestingFeatureInteractionsBriandNsgaForConcreteScenes}
\begin{bchapter}
\bauthor{\bsnm{Abdessalem}, \binits{R.B.}},
\bauthor{\bsnm{Panichella}, \binits{A.}},
\bauthor{\bsnm{Nejati}, \binits{S.}},
\bauthor{\bsnm{Briand}, \binits{L.C.}},
\bauthor{\bsnm{Stifter}, \binits{T.}}:
\bctitle{Testing autonomous cars for feature interaction failures using
  many-objective search}.
In: \bbtitle{ASE 2018 - Proceedings of the 33rd ACM/IEEE Int'l Conf. on
  Automated Software Engineering},
vol. \bseriesno{18}.
\bconflocation{New York, New York, USA},
pp. \bfpage{143}--\blpage{154}
(\byear{2018})
\end{bchapter}
\endbibitem

\bibitem[\protect\citeauthoryear{Babikian et~al.}{2022}]{Babikian2021dReal}
\begin{barticle}
\bauthor{\bsnm{Babikian}, \binits{A.A.}},
\bauthor{\bsnm{Semer{\'{a}}th}, \binits{O.}},
\bauthor{\bsnm{Li}, \binits{A.}},
\bauthor{\bsnm{Marussy}, \binits{K.}},
\bauthor{\bsnm{Varr{\'{o}}}, \binits{D.}}:
\batitle{Automated generation of consistent models using qualitative
  abstractions and exploration strategies}.
\bjtitle{Softw. Syst. Model.}
\bvolume{21}(\bissue{5}),
\bfpage{1763}--\blpage{1787}
(\byear{2022})
\end{barticle}
\endbibitem

\bibitem[\protect\citeauthoryear{Babikian
  et~al.}{2024}]{babikianConcretizationAbstractTraffic2024}
\begin{barticle}
\bauthor{\bsnm{Babikian}, \binits{A.A.}},
\bauthor{\bsnm{Semer{\'a}th}, \binits{O.}},
\bauthor{\bsnm{Varr{\'o}}, \binits{D.}}:
\batitle{Concretization of {{Abstract Traffic Scene Specifications Using
  Metaheuristic Search}}}.
\bjtitle{IEEE Transactions on Software Engineering}
\bvolume{50}(\bissue{1}),
\bfpage{48}--\blpage{68}
(\byear{2024})
\end{barticle}
\endbibitem

\bibitem[\protect\citeauthoryear{Alexander
  et~al.}{2015}]{Alexander2015SituationCoverage}
\begin{bbook}
\bauthor{\bsnm{Alexander}, \binits{R.}},
\bauthor{\bsnm{Hawkins}, \binits{H.}},
\bauthor{\bsnm{Rae}, \binits{D.}}:
\bbtitle{Situation Coverage – a Coverage Criterion for Testing Autonomous
  Robots},
pp. \bfpage{1}--\blpage{20}
(\byear{2015})
\end{bbook}
\endbibitem

\bibitem[\protect\citeauthoryear{Ulbrich et~al.}{2015}]{Ulbrich2015Defining}
\begin{bchapter}
\bauthor{\bsnm{Ulbrich}, \binits{S.}},
\bauthor{\bsnm{Menzel}, \binits{T.}},
\bauthor{\bsnm{Reschka}, \binits{A.}},
\bauthor{\bsnm{Schuldt}, \binits{F.}},
\bauthor{\bsnm{Maurer}, \binits{M.}}:
\bctitle{Defining and substantiating the terms scene, situation, and scenario
  for automated driving}.
In: \bbtitle{IEEE Conference on Intelligent Transportation Systems,
  Proceedings, ITSC},
vol. \bseriesno{2015-Octob},
pp. \bfpage{982}--\blpage{988}
(\byear{2015})
\end{bchapter}
\endbibitem

\bibitem[\protect\citeauthoryear{Abdessalem
  et~al.}{2018}]{Abdessalem2018TestingVisionBased}
\begin{bchapter}
\bauthor{\bsnm{Abdessalem}, \binits{R.B.}},
\bauthor{\bsnm{Nejati}, \binits{S.}},
\bauthor{\bsnm{Briand}, \binits{L.C.}},
\bauthor{\bsnm{Stifter}, \binits{T.}}:
\bctitle{Testing vision-based control systems using learnable evolutionary
  algorithms}.
In: \bbtitle{Proceedings - Int'l Conf. on Software Engineering},
vol. \bseriesno{11},
pp. \bfpage{1016}--\blpage{1026}
(\byear{2018})
\end{bchapter}
\endbibitem

\bibitem[\protect\citeauthoryear{Haq et~al.}{2023}]{Haq2023Reinforcement}
\begin{botherref}
\oauthor{\bsnm{Haq}, \binits{F.U.}},
\oauthor{\bsnm{Shin}, \binits{D.}},
\oauthor{\bsnm{Briand}, \binits{L.C.}}:
Many-objective reinforcement learning for online testing of dnn-enabled
  systems.
2023 IEEE/ACM 45th International Conference on Software Engineering (ICSE),
1814--1826
(2023)
\end{botherref}
\endbibitem

\bibitem[\protect\citeauthoryear{Zhong et~al.}{2021}]{Zhong2021Fuzzing}
\begin{botherref}
\oauthor{\bsnm{Zhong}, \binits{Z.}},
\oauthor{\bsnm{Kaiser}, \binits{G.}},
\oauthor{\bsnm{Ray}, \binits{B.}}:
Neural network guided evolutionary fuzzing for finding traffic violations of
  autonomous vehicles.
IEEE Transactions on Software Engineering
(2021)
\end{botherref}
\endbibitem

\bibitem[\protect\citeauthoryear{Calo
  et~al.}{2020}]{calo2020GeneratingAvoidableCollision}
\begin{bchapter}
\bauthor{\bsnm{Calo}, \binits{A.}},
\bauthor{\bsnm{Arcaini}, \binits{P.}},
\bauthor{\bsnm{Ali}, \binits{S.}},
\bauthor{\bsnm{Hauer}, \binits{F.}},
\bauthor{\bsnm{Ishikawa}, \binits{F.}}:
\bctitle{Generating avoidable collision scenarios for testing autonomous
  driving systems}.
In: \bbtitle{Proceedings - 2020 IEEE 13th Int'l Conf. on Software Testing,
  Verification and Validation, ICST 2020},
pp. \bfpage{375}--\blpage{386}
(\byear{2020})
\end{bchapter}
\endbibitem

\bibitem[\protect\citeauthoryear{Wu et~al.}{2021}]{Wu2021}
\begin{botherref}
\oauthor{\bsnm{Wu}, \binits{S.}},
\oauthor{\bsnm{Wang}, \binits{H.}},
\oauthor{\bsnm{Yu}, \binits{W.}},
\oauthor{\bsnm{Yang}, \binits{K.}},
\oauthor{\bsnm{Cao}, \binits{D.}},
\oauthor{\bsnm{Wang}, \binits{F.}}:
A new {SOTIF} scenario hierarchy and its critical test case generation based on
  potential risk assessment.
IEEE 1st Int'l Conf. on Digital Twins and Parallel Intelligence, DTPI 2021,
399--409
(2021)
\end{botherref}
\endbibitem

\bibitem[\protect\citeauthoryear{Fremont
  et~al.}{2019}]{Fremont2019ScenicLanguage}
\begin{bchapter}
\bauthor{\bsnm{Fremont}, \binits{D.J.}},
\bauthor{\bsnm{Yue}, \binits{X.}},
\bauthor{\bsnm{Dreossi}, \binits{T.}},
\bauthor{\bsnm{Sangiovanni-Vincentelli}, \binits{A.L.}},
\bauthor{\bsnm{Ghosh}, \binits{S.}},
\bauthor{\bsnm{Seshia}, \binits{S.A.}}:
\bctitle{Scenic: A language for scenario specification and scene generation}.
In: \bbtitle{Proceedings of the ACM SIGPLAN Conference on Programming Language
  Design and Implementation (PLDI)},
pp. \bfpage{63}--\blpage{78}
(\byear{2019})
\end{bchapter}
\endbibitem

\bibitem[\protect\citeauthoryear{Haq et~al.}{2022}]{Haq2022}
\begin{barticle}
\bauthor{\bsnm{Haq}, \binits{F.U.}},
\bauthor{\bsnm{Shin}, \binits{D.}},
\bauthor{\bsnm{Briand}, \binits{L.}}:
\batitle{Efficient online testing for dnn-enabled systems using
  surrogate-assisted and many-objective optimization}.
\bjtitle{Proceedings - International Conference on Software Engineering}
\bvolume{2022-May},
\bfpage{811}--\blpage{822}
(\byear{2022})
\end{barticle}
\endbibitem

\bibitem[\protect\citeauthoryear{Riccio and Tonella}{2020}]{Riccio2020}
\begin{botherref}
\oauthor{\bsnm{Riccio}, \binits{V.}},
\oauthor{\bsnm{Tonella}, \binits{P.}}:
Model-based exploration of the frontier of behaviours for deep learning system
  testing.
ESEC/FSE 2020 - Proceedings of the 28th ACM Joint Meeting European Software
  Engineering Conference and Symposium on the Foundations of Software
  Engineering,
876--888
(2020)
\end{botherref}
\endbibitem

\bibitem[\protect\citeauthoryear{Menzel
  et~al.}{2018}]{Menzel2018ScenariosForDevelopment}
\begin{bchapter}
\bauthor{\bsnm{Menzel}, \binits{T.}},
\bauthor{\bsnm{Bagschik}, \binits{G.}},
\bauthor{\bsnm{Maurer}, \binits{M.}}:
\bctitle{Scenarios for development, test and validation of automated vehicles}.
In: \bbtitle{IEEE Intelligent Vehicles Symposium, Proceedings},
vol. \bseriesno{2018-June},
pp. \bfpage{1821}--\blpage{1827}
(\byear{2018})
\end{bchapter}
\endbibitem

\bibitem[\protect\citeauthoryear{Dosovitskiy
  et~al.}{2017}]{Dosovitskiy2017CarlaSimulator}
\begin{bchapter}
\bauthor{\bsnm{Dosovitskiy}, \binits{A.}},
\bauthor{\bsnm{Ros}, \binits{G.}},
\bauthor{\bsnm{Codevilla}, \binits{F.}},
\bauthor{\bsnm{Lopez}, \binits{A.}},
\bauthor{\bsnm{Koltun}, \binits{V.}}:
\bctitle{Carla: An open urban driving simulator}.
In: \bbtitle{Proceedings of the 1st Annual Conference on Robot Learning},
pp. \bfpage{1}--\blpage{16}
(\byear{2017})
\end{bchapter}
\endbibitem

\bibitem[\protect\citeauthoryear{Prakash et~al.}{2021}]{Prakash2021CVPR}
\begin{bchapter}
\bauthor{\bsnm{Prakash}, \binits{A.}},
\bauthor{\bsnm{Chitta}, \binits{K.}},
\bauthor{\bsnm{Geiger}, \binits{A.}}:
\bctitle{Multi-modal fusion transformer for end-to-end autonomous driving}.
In: \bbtitle{Conference on Computer Vision and Pattern Recognition (CVPR)}
(\byear{2021})
\end{bchapter}
\endbibitem

\bibitem[\protect\citeauthoryear{Chitta et~al.}{2023}]{Chitta2023PAMI}
\begin{botherref}
\oauthor{\bsnm{Chitta}, \binits{K.}},
\oauthor{\bsnm{Prakash}, \binits{A.}},
\oauthor{\bsnm{Jaeger}, \binits{B.}},
\oauthor{\bsnm{Yu}, \binits{Z.}},
\oauthor{\bsnm{Renz}, \binits{K.}},
\oauthor{\bsnm{Geiger}, \binits{A.}}:
Transfuser: Imitation with transformer-based sensor fusion for autonomous
  driving.
Pattern Analysis and Machine Intelligence (PAMI)
(2023)
\end{botherref}
\endbibitem

\bibitem[\protect\citeauthoryear{Abdessalem
  et~al.}{2016}]{Abdessalem2016TestingADAS}
\begin{botherref}
\oauthor{\bsnm{Abdessalem}, \binits{R.B.}},
\oauthor{\bsnm{Nejati}, \binits{S.}},
\oauthor{\bsnm{Briand}, \binits{L.C.}},
\oauthor{\bsnm{Stifter}, \binits{T.}}:
Testing advanced driver assistance systems using multi-objective search and
  neural networks.
ASE 2016 - Proceedings of the 31st IEEE/ACM Int'l Conf. on Automated Software
  Engineering,
63--74
(2016)
\end{botherref}
\endbibitem

\bibitem[\protect\citeauthoryear{Klischat and Althoff}{2019}]{Klischat2019}
\begin{bchapter}
\bauthor{\bsnm{Klischat}, \binits{M.}},
\bauthor{\bsnm{Althoff}, \binits{M.}}:
\bctitle{Generating critical test scenarios for automated vehicles with
  evolutionary algorithms}.
In: \bbtitle{IEEE Intelligent Vehicles Symposium, Proceedings},
vol. \bseriesno{2019-June},
pp. \bfpage{2352}--\blpage{2358}
(\byear{2019})
\end{bchapter}
\endbibitem

\bibitem[\protect\citeauthoryear{Fisher}{1992}]{Fisher1992FishersExactTest}
\begin{botherref}
\oauthor{\bsnm{Fisher}, \binits{R.A.}}:
Statistical methods for research workers.
Breakthroughs in statistics,
66--70
(1992)
\end{botherref}
\endbibitem

\bibitem[\protect\citeauthoryear{Haddock et~al.}{1998}]{Haddock1998OddsRatio}
\begin{barticle}
\bauthor{\bsnm{Haddock}, \binits{C.K.}},
\bauthor{\bsnm{Rindskopf}, \binits{D.}},
\bauthor{\bsnm{Shadish}, \binits{W.R.}}:
\batitle{Using odds ratios as effect sizes for meta-analysis of dichotomous
  data: A primer on methods and issues}.
\bjtitle{Psychological Methods}
\bvolume{3}(\bissue{3}),
\bfpage{339}--\blpage{353}
(\byear{1998})
\end{barticle}
\endbibitem

\bibitem[\protect\citeauthoryear{Arcuri and Briand}{2011}]{Arcuri2011}
\begin{botherref}
\oauthor{\bsnm{Arcuri}, \binits{A.}},
\oauthor{\bsnm{Briand}, \binits{L.}}:
A practical guide for using statistical tests to assess randomized algorithms
  in software engineering.
Proceedings - Int'l Conf. on Software Engineering,
1--10
(2011)
\end{botherref}
\endbibitem

\bibitem[\protect\citeauthoryear{Lu
  et~al.}{2023}]{luLearningConfigurationsOperating2023}
\begin{barticle}
\bauthor{\bsnm{Lu}, \binits{C.}},
\bauthor{\bsnm{Shi}, \binits{Y.}},
\bauthor{\bsnm{Zhang}, \binits{H.}},
\bauthor{\bsnm{Zhang}, \binits{M.}},
\bauthor{\bsnm{Wang}, \binits{T.}},
\bauthor{\bsnm{Yue}, \binits{T.}},
\bauthor{\bsnm{Ali}, \binits{S.}}:
\batitle{Learning {{Configurations}} of {{Operating Environment}} of
  {{Autonomous Vehicles}} to {{Maximize}} their {{Collisions}}}.
\bjtitle{IEEE Transactions on Software Engineering}
\bvolume{49}(\bissue{1}),
\bfpage{384}--\blpage{402}
(\byear{2023})
\end{barticle}
\endbibitem

\bibitem[\protect\citeauthoryear{Steimle et~al.}{2021}]{steimle2021}
\begin{barticle}
\bauthor{\bsnm{Steimle}, \binits{M.}},
\bauthor{\bsnm{Menzel}, \binits{T.}},
\bauthor{\bsnm{Maurer}, \binits{M.}}:
\batitle{Toward a consistent taxonomy for scenario-based development and test
  approaches for automated vehicles: A proposal for a structuring framework, a
  basic vocabulary, and its application}.
\bjtitle{IEEE Access}
\bvolume{9},
\bfpage{147828}--\blpage{147854}
(\byear{2021})
{\href{https://arxiv.org/abs/2104.09097v3}{{arXiv:2104.09097v3}}}
\end{barticle}
\endbibitem

\bibitem[\protect\citeauthoryear{Schuldt et~al.}{2018}]{Schuldt2018}
\begin{bchapter}
\bauthor{\bsnm{Schuldt}, \binits{F.}},
\bauthor{\bsnm{Reschka}, \binits{A.}},
\bauthor{\bsnm{Maurer}, \binits{M.}}:
\bctitle{A method for an efficient, systematic test case generation for
  advanced driver assistance systems in virtual environments}.
In: \bbtitle{Automotive Systems Engineering II},
pp. \bfpage{147}--\blpage{175}
(\byear{2018})
\end{bchapter}
\endbibitem

\bibitem[\protect\citeauthoryear{Bagschik et~al.}{2018}]{Bagschik2018}
\begin{bchapter}
\bauthor{\bsnm{Bagschik}, \binits{G.}},
\bauthor{\bsnm{Menzel}, \binits{T.}},
\bauthor{\bsnm{Maurer}, \binits{M.}}:
\bctitle{Ontology based scene creation for the development of automated
  vehicles}.
In: \bbtitle{IEEE Intelligent Vehicles Symposium},
pp. \bfpage{1813}--\blpage{1820}
(\byear{2018})
\end{bchapter}
\endbibitem

\bibitem[\protect\citeauthoryear{Scholtes et~al.}{2021}]{Scholtes2021}
\begin{barticle}
\bauthor{\bsnm{Scholtes}, \binits{M.}},
\bauthor{\bsnm{Westhofen}, \binits{L.}},
\bauthor{\bsnm{Turner}, \binits{L.R.}},
\bauthor{\bsnm{Lotto}, \binits{K.}},
\bauthor{\bsnm{Schuldes}, \binits{M.}},
\bauthor{\bsnm{Weber}, \binits{H.}},
\bauthor{\bsnm{Wagener}, \binits{N.}},
\bauthor{\bsnm{Neurohr}, \binits{C.}},
\bauthor{\bsnm{Bollmann}, \binits{M.H.}},
\bauthor{\bsnm{Kortke}, \binits{F.}},
\bauthor{\bsnm{Hiller}, \binits{J.}},
\bauthor{\bsnm{Hoss}, \binits{M.}},
\bauthor{\bsnm{Bock}, \binits{J.}},
\bauthor{\bsnm{Eckstein}, \binits{L.}}:
\batitle{6-layer model for a structured description and categorization of urban
  traffic and environment}.
\bjtitle{IEEE Access}
\bvolume{9},
\bfpage{59131}--\blpage{59147}
(\byear{2021})
{\href{https://arxiv.org/abs/2012.06319}{{arXiv:2012.06319}}}
\end{barticle}
\endbibitem

\bibitem[\protect\citeauthoryear{Urbieta et~al.}{2021}]{Urbieta2021}
\begin{barticle}
\bauthor{\bsnm{Urbieta}, \binits{I.}},
\bauthor{\bsnm{Nieto}, \binits{M.}},
\bauthor{\bsnm{García}, \binits{M.}},
\bauthor{\bsnm{Otaegui}, \binits{O.}},
\bauthor{\bsnm{Clavijo}, \binits{M.}},
\bauthor{\bsnm{Jiménez}, \binits{F.}},
\bauthor{\bsnm{Naranjo}, \binits{J.E.}}:
\batitle{Design and implementation of an ontology for semantic labeling and
  testing: Automotive global ontology (ago)}.
\bjtitle{Applied Sciences}
\bvolume{11}(\bissue{17}),
\bfpage{7782}
(\byear{2021})
\end{barticle}
\endbibitem

\bibitem[\protect\citeauthoryear{Geyer et~al.}{2014}]{Geyer2014}
\begin{barticle}
\bauthor{\bsnm{Geyer}, \binits{S.}},
\bauthor{\bsnm{Baltzer}, \binits{M.}},
\bauthor{\bsnm{Franz}, \binits{B.}},
\bauthor{\bsnm{Hakuli}, \binits{S.}},
\bauthor{\bsnm{Kauer}, \binits{M.}},
\bauthor{\bsnm{Kienle}, \binits{M.}},
\bauthor{\bsnm{Meier}, \binits{S.}},
\bauthor{\bsnm{Weigerber}, \binits{T.}},
\bauthor{\bsnm{Bengler}, \binits{K.}},
\bauthor{\bsnm{Bruder}, \binits{R.}},
\bauthor{\bsnm{Flemisch}, \binits{F.}},
\bauthor{\bsnm{Winner}, \binits{H.}}:
\batitle{Concept and development of a unified ontology for generating test and
  use-case catalogues for assisted and automated vehicle guidance}.
\bjtitle{IET Intelligent Transport Systems}
\bvolume{8}(\bissue{3}),
\bfpage{183}--\blpage{189}
(\byear{2014})
\end{barticle}
\endbibitem

\bibitem[\protect\citeauthoryear{Klueck et~al.}{2018}]{Klueck2018}
\begin{bchapter}
\bauthor{\bsnm{Klueck}, \binits{F.}},
\bauthor{\bsnm{Li}, \binits{Y.}},
\bauthor{\bsnm{Nica}, \binits{M.}},
\bauthor{\bsnm{Tao}, \binits{J.}},
\bauthor{\bsnm{Wotawa}, \binits{F.}}:
\bctitle{Using ontologies for test suites generation for automated and
  autonomous driving functions}.
In: \bbtitle{Proceedings - 29th IEEE International Symposium on Software
  Reliability Engineering Workshops, ISSREW 2018},
pp. \bfpage{118}--\blpage{123}
(\byear{2018})
\end{bchapter}
\endbibitem

\bibitem[\protect\citeauthoryear{Queiroz
  et~al.}{2019}]{Queiroz2019GeoScenarioAn}
\begin{bchapter}
\bauthor{\bsnm{Queiroz}, \binits{R.}},
\bauthor{\bsnm{Berger}, \binits{T.}},
\bauthor{\bsnm{Czarnecki}, \binits{K.}}:
\bctitle{{GeoScenario}: An open {DSL} for autonomous driving scenario
  representation}.
In: \bbtitle{IEEE Intelligent Vehicles Symposium},
pp. \bfpage{287}--\blpage{294}
(\byear{2019})
\end{bchapter}
\endbibitem

\bibitem[\protect\citeauthoryear{Schütt et~al.}{2020}]{Schutt2020}
\begin{bchapter}
\bauthor{\bsnm{Schütt}, \binits{B.}},
\bauthor{\bsnm{Braun}, \binits{T.}},
\bauthor{\bsnm{Otten}, \binits{S.}},
\bauthor{\bsnm{Sax}, \binits{E.}}:
\bctitle{{SceML}: A graphical modeling framework for scenario-based testing of
  autonomous vehicles}.
In: \bbtitle{23rd ACM/IEEE Int'l Conf. on Model Driven Engineering Languages
  and Systems},
pp. \bfpage{114}--\blpage{120}
(\byear{2020})
\end{bchapter}
\endbibitem

\bibitem[\protect\citeauthoryear{Hempen et~al.}{2017}]{Hempen2017}
\begin{barticle}
\bauthor{\bsnm{Hempen}, \binits{T.}},
\bauthor{\bsnm{Biank}, \binits{S.}},
\bauthor{\bsnm{Huber}, \binits{W.}},
\bauthor{\bsnm{Diedrich}, \binits{C.}}:
\batitle{Model based generation of driving scenarios}.
\bjtitle{Lecture Notes of the Institute for Computer Sciences,
  Social-Informatics and Telecommunications Engineering, LNICST}
\bvolume{222},
\bfpage{153}--\blpage{163}
(\byear{2017})
\end{barticle}
\endbibitem

\bibitem[\protect\citeauthoryear{Bach
  et~al.}{2016}]{Bach2016ModelBasedScenario}
\begin{bchapter}
\bauthor{\bsnm{Bach}, \binits{J.}},
\bauthor{\bsnm{Otten}, \binits{S.}},
\bauthor{\bsnm{Sax}, \binits{E.}}:
\bctitle{Model based scenario specification for development and test of
  automated driving functions}.
In: \bbtitle{IEEE Intelligent Vehicles Symposium},
pp. \bfpage{1149}--\blpage{1155}
(\byear{2016})
\end{bchapter}
\endbibitem

\bibitem[\protect\citeauthoryear{Ries et~al.}{2019}]{Ries2019}
\begin{botherref}
\oauthor{\bsnm{Ries}, \binits{L.}},
\oauthor{\bsnm{Langner}, \binits{J.}},
\oauthor{\bsnm{Otten}, \binits{S.}},
\oauthor{\bsnm{Bach}, \binits{J.}},
\oauthor{\bsnm{Sax}, \binits{E.}}:
A driving scenario representation for scalable real-data analytics with neural
  networks.
IEEE Intelligent Vehicles Symposium,
2215--2222
(2019)
\end{botherref}
\endbibitem

\bibitem[\protect\citeauthoryear{Panichella et~al.}{2015}]{Panichella2015}
\begin{botherref}
\oauthor{\bsnm{Panichella}, \binits{A.}},
\oauthor{\bsnm{Kifetew}, \binits{F.M.}},
\oauthor{\bsnm{Tonella}, \binits{P.}}:
Reformulating branch coverage as a many-objective optimization problem.
2015 IEEE 8th International Conference on Software Testing, Verification and
  Validation, ICST 2015 - Proceedings
(2015)
\end{botherref}
\endbibitem

\bibitem[\protect\citeauthoryear{Fan et~al.}{2017}]{Fan2017}
\begin{botherref}
\oauthor{\bsnm{Fan}, \binits{Z.}},
\oauthor{\bsnm{Yi}, \binits{F.}},
\oauthor{\bsnm{Li}, \binits{W.}},
\oauthor{\bsnm{Lu}, \binits{J.}},
\oauthor{\bsnm{Cai}, \binits{X.}},
\oauthor{\bsnm{Wei}, \binits{C.}}:
A comparative study of constrained multi-objective evolutionary algorithms on
  constrained multi-objective optimization problems.
IEEE Congress on Evolutionary Computation, CEC 2017,
209--216
(2017)
\end{botherref}
\endbibitem

\bibitem[\protect\citeauthoryear{Majumdar
  et~al.}{2021}]{Majumdar2021ParacsmJournal}
\begin{bchapter}
\bauthor{\bsnm{Majumdar}, \binits{R.}},
\bauthor{\bsnm{Mathur}, \binits{A.}},
\bauthor{\bsnm{Pirron}, \binits{M.}},
\bauthor{\bsnm{Stegner}, \binits{L.}},
\bauthor{\bsnm{Zufferey}, \binits{D.}}:
\bctitle{Paracosm: A test framework for autonomous driving simulations}.
In: \bbtitle{Fundamental Approaches to Software Engineering},
pp. \bfpage{172}--\blpage{195}
(\byear{2021})
\end{bchapter}
\endbibitem

\bibitem[\protect\citeauthoryear{Rocklage
  et~al.}{2018}]{Rocklage2018AutomatedScenarioGeneration}
\begin{bchapter}
\bauthor{\bsnm{Rocklage}, \binits{E.}},
\bauthor{\bsnm{Kraft}, \binits{H.}},
\bauthor{\bsnm{Karatas}, \binits{A.}},
\bauthor{\bsnm{Seewig}, \binits{J.}}:
\bctitle{Automated scenario generation for regression testing of autonomous
  vehicles}.
In: \bbtitle{IEEE Conference on Intelligent Transportation Systems,
  Proceedings, ITSC},
vol. \bseriesno{2018-March},
pp. \bfpage{476}--\blpage{483}
(\byear{2018})
\end{bchapter}
\endbibitem

\bibitem[\protect\citeauthoryear{O'Kelly
  et~al.}{2018}]{OKelly2018ScalableEndToEnd}
\begin{bchapter}
\bauthor{\bsnm{O'Kelly}, \binits{M.}},
\bauthor{\bsnm{Duchi}, \binits{J.}},
\bauthor{\bsnm{Sinha}, \binits{A.}},
\bauthor{\bsnm{Namkoong}, \binits{H.}},
\bauthor{\bsnm{Tedrake}, \binits{R.}}:
\bctitle{Scalable end-to-end autonomous vehicle testing via rare-event
  simulation}.
In: \bbtitle{Advances in Neural Information Processing Systems},
vol. \bseriesno{2018-Decem},
pp. \bfpage{9827}--\blpage{9838}
(\byear{2018})
\end{bchapter}
\endbibitem

\bibitem[\protect\citeauthoryear{Althoff and Lutz}{2018}]{Althoff2018}
\begin{bchapter}
\bauthor{\bsnm{Althoff}, \binits{M.}},
\bauthor{\bsnm{Lutz}, \binits{S.}}:
\bctitle{Automatic generation of safety-critical test scenarios for collision
  avoidance of road vehicles}.
In: \bbtitle{IEEE Intelligent Vehicles Symposium},
pp. \bfpage{1326}--\blpage{1333}
(\byear{2018})
\end{bchapter}
\endbibitem

\end{thebibliography}







\end{document}